\documentclass[a4paper,hidelinks, 12pt]{article}
\usepackage[utf8]{inputenc}
\usepackage{a4wide}
\usepackage{graphics}
\usepackage{amsmath}
\usepackage{amsfonts}
\usepackage{amssymb}
\usepackage{subfigure}
\usepackage{cite}
\usepackage[table]{xcolor}
\usepackage{soul}
\usepackage[toc,page]{appendix}
\usepackage{mathtools}
\usepackage{hyperref}
\usepackage{enumitem}
\usepackage{threeparttable}
\usepackage{etoolbox} 
\usepackage{longtable}
\usepackage{multirow}
\usepackage{threeparttablex}
\usepackage{booktabs}
\usepackage{pdflscape}
\usepackage{float}
\usepackage{makecell}
\usepackage{hhline}

\allowdisplaybreaks
\NewCommandCopy{\originalsubsection}{\subsection}
\RenewDocumentCommand{\subsection}{sd()O{#4}m}{%
  \IfBooleanTF{#1}
    {\originalsubsection*{#4}}%
    {%
     \IfNoValueTF{#2}
       {
        \renewcommand{\thesubsection}{\thesection.\arabic{subsection}}%
       }
       {
        \addtocounter{subsection}{-1}%
        \renewcommand{\thesubsection}{\thesection#2}%
       }
     \originalsubsection[#3]{#4}%
    }%
}

\numberwithin{equation}{section}

\makeatletter
\@addtoreset{equation}{section}
\let\TPT@hookin\@gobble
\let\TPT@hookarg\@gobble
\newcommand{\labeltext}[2]{%
  \@bsphack
  \csname phantomsection\endcsname
  \def\@currentlabel{#1}{\label{#2}}%
  \@esphack
}
\makeatother

\title{Draft on GOOFy 3HDMs}
\date{\today}

\begin{document}

\begin{titlepage}
\begin{center}

{\large \bf {GOOFy-compatible 3HDMs and beyond}}

\vskip 1cm

I. de Medeiros Varzielas$^{a,}$\footnote{E-mail:  ivo.de@udo.edu } and
A. Kun\v cinas$^{b,}$\footnote{E-mail: Anton.Kuncinas@uv.es}

\vspace{1.0cm}

$^{a}$Centro de F\'isica Te\'orica de Part\'iculas, CFTP, Departamento de F\'\i sica,\\ Instituto Superior T\'ecnico, Universidade de Lisboa,\\
Avenida Rovisco Pais nr. 1, 1049-001 Lisboa, Portugal

$^{b}$Instituto de Física Corpuscular (IFIC), CSIC‐Universitat de València,\\ Parc Científic UV, c/ Catedrático José Beltrán, 2, E-46980 Paterna (València), Spain

\end{center}

\vskip 2cm

\begin{abstract}

\noindent

Beyond conventional Higgs-family and general CP transformations, one may also consider a broader class of non-standard ``GOOFy'' transformations, in which the scalar fields and their conjugates are assigned related but inequivalent transformations. Although these transformations are not symmetries of a full Lagrangian in the conventional sense, some nevertheless stabilise the scalar potential. Their impact on the quadratic sector has not yet been systematically classified. We develop a sign-orbit technique to determine the bilinear structures compatible with these transformations. Applied to three-Higgs-doublet models, and formulated so as to extend to larger multi-Higgs sectors, the method shows that a non-vanishing GOOFy-compatible quadratic sector exists only when two independent conditions are met: the underlying group admits a non-trivial sign character, and that character is realised within the tensor product of the scalar representation and its conjugate. As a key consequence, we demonstrate that the renormalisation-group stability of the $r_0 \to -r_0$ manifold in isolation is not a generic feature of multi-Higgs potentials, but rather a direct consequence of special algebraic properties of $SU(2)$. These results provide a classification of the admissible GOOFy quadratic structures and their invariant bilinear subspaces, highlighting the algebraic obstructions to their renormalisation-group stability.

\end{abstract}

\end{titlepage}

 \tableofcontents
 
 \newpage

\section{Introduction}

The Standard Model (SM) of particle physics relies on a single $SU(2)_L \times U(1)_Y$ scalar doublet, which is responsible for spontaneous electroweak symmetry breaking and the generation of the masses of the fundamental fermions. While the discovery of a 125 GeV Higgs boson~\cite{ATLAS:2012yve,CMS:2012qbp} confirmed the basic premise of this mechanism, the minimal scalar sector offers no structural explanations for some of the most profound puzzles in high-energy physics, including the origins of neutrino masses, the matter-antimatter asymmetry, the nature of dark matter, and the extreme fine-tuning required to stabilise the electroweak scale against the high-energy quantum corrections. Consequently, extending the scalar sector by introducing additional scalar doublets constitutes one of the most compelling directions for physics beyond the SM.

Multi-Higgs-doublet models (NHDMs), frameworks containing $N$ copies of the $SU(2)_L$ scalar doublet, provide a rich phenomenology that accommodates spontaneous CP violation, additional flavour dynamics, and dark matter candidates. However, this structural flexibility comes at the cost of additional parameters. The generic scalar potential of the two-Higgs-doublet model (2HDM) contains 14 real parameters (11 independent ones), and in the three-Higgs-doublet model (3HDM) the parameter space is characterised by 54 parameters (46 independent ones). The predictiveness of these models is quickly lost due to such a high number of unconstrained parameters~\cite{Olaussen:2010aq}, introducing $N_\text{phys} = (N^4 + N^2 + 2)/2$ physical parameters for $N$ scalar doublets.

To render these models predictive and phenomenologically viable, it is standard practice to impose global symmetry groups on the scalar potential. Historically, the imposition of symmetries has typically been restricted to two well-defined symmetry transformations: Higgs Family (HF) transformations, which dictate unitary mixing transformations among the scalar doublets, and General CP (GCP) transformations, which combine unitary mixing with complex conjugation. Over the past decade, systematic classifications of realisable HF and GCP symmetries have been achieved in 2HDMs~\cite{Ivanov:2005hg,Ivanov:2006yq,Gerard:2007kn,Ivanov:2007de,Ferreira:2009wh,Ferreira:2010yh,Battye:2011jj,Pilaftsis:2011ed,Haber:2018iwr,Bento:2020jei,Ferreira:2020ana, Ferreira:2022gjh,Solberg:2025ybf}, 3HDMs~\cite{Ferreira:2008zy,Ivanov:2011ae,Ivanov:2012ry,Ivanov:2012fp,Keus:2013hya,Ivanov:2014doa,Ivanov:2015mwl,Pilaftsis:2016erj,deMedeirosVarzielas:2019rrp,KunMT,Darvishi:2019dbh,Kuncinas:2020wrn, Kuncinas:2024zjq, Doring:2024kdg,Kuncinas:2025uty}, and are ongoing in 4HDMs~\cite{Shao:2023oxt,Shao:2024ibu}.

Recently, however, conventional paradigms governing symmetry transformations have been challenged by the introduction of an entirely new class of transformations. Originally identified in the 2HDM by Ferreira, Grzadkowski, Ogreid and Osland~\cite{Ferreira:2023dke}, these transformations enforce highly specific parameter relations within the scalar potential that are stable under renormalisation group (RG) equations at least to order three, yet they do not correspond to any known HF or GCP symmetries. Dubbed ``GOOFy" transformations, these operations possess the distinctive property of non-trivially transforming the canonical gauge-kinetic terms of the Lagrangian, enforcing a kinetic sign flip. The theoretical investigation of GOOFy transformations is an actively developing field; foundational aspects are discussed in Refs.~\cite{Haber:2025cbb,Trautner:2025yxz,Ferreira:2025ate}, whilst their implications for the Yukawa sector are detailed in Refs.~\cite{deBoer:2025jhc,Trautner:2025prm,Grzadkowski:2026gkx,Ferreira:2026ccg}.

A recent analysis~\cite{Kuncinas:2025uty} of the 3HDM has incorporated GOOFy transformations into the classification of scalar potentials, identifying the invariant structures of the quartic sector. However, classification of the quadratic scalar sector was abandoned in this preliminary study. As we shall argue, the quadratic part of the potential is sensitive to the specific group representations chosen for the fields relative to their complex conjugates. 

This report provides a theoretical framework to classify and analyse how the quadratic terms behave under the action of GOOFy transformations in 3HDMs. First of all, in Section~\ref{Sec:Framework} we review realisable HF and GCP symmetry groups and comment on some properties of the quadratic couplings. Next, in Section~\ref{Sec:GOOFY_transformations} we define the GOOFy transformations and outline their key characteristics; specifically, we question whether these can be extended to NHDMs or if it might be just an accident of the 2HDM. Then, in Section~\ref{Sec:GOOFY_admissible} we analyse GOOFy transformations from a group-theoretical and representation-theoretical perspective, formulating some ideas based on the representation mismatch. Having outlined the theoretical part in the previous sections, we move on to Section~\ref{Sec:Realisable_GOOFy}, where we identify a set of different 3HDM potentials within the GOOFy framework and consider their physical spectra and RG running.

\section{Framework}\label{Sec:Framework}

Before discussing GOOFy transformations, we set up notation. The most general renormalisable NHDM potential is constructed from the gauge-invariant bilinear covariants
\begin{equation}
h_{ij} \equiv h_i^\dagger h_j.
\end{equation}
Note that $h_i^\dagger h_j = h_j^\mathrm{T} h_i^\ast$.

The scalar potential can be written as
\begin{equation}\label{Eq:V_NHDM_gen}
V= \mu_{ij}^2 h_{ij} + \lambda_{ijkl} h_{ij}h_{kl}.
\end{equation}
We denote the tensors of quadratic and quartic coefficients by $Y_{ij}\equiv \mu_{ij}^2$ and $Z_{ijkl}\equiv \lambda_{ijkl}$. In the following, we will refer interchangeably to the quadratic and quartic coefficient tensors as $Y$ and $Z$, with their indices suppressed whenever no ambiguity arises.

The potential is required to be Hermitian, which enforces two conditions: $\mu_{ij}^2 = (\mu_{ji}^2)^\ast$ and $\lambda_{ijkl} = \lambda_{klij}= \lambda_{klij}^\ast$. 

The physical observables of the NHDM are invariant under a unitary basis change, $h_i \mapsto U_{ij} h_j$. Because the scalar fields can be arbitrarily rotated, identifying the full symmetry group of a potential is complicated by basis dependence, which often renders an underlying symmetry non-manifest.

To systematically tackle symmetries, one must identify realisable groups---that is, symmetry groups whose imposition on the potential does not accidentally induce a larger symmetry. In the 3HDM, any transformation acting on the scalar potential, relies on identifying the irreducible representations of finite subgroups of $\mathcal{U} \in U(3)$, whether it is:
\begin{itemize}
\item an HF transformation (these are block-diagonal),
\begin{equation}\label{Eq:H_to_HF}
\begin{pmatrix} h_i \\ h_i^\ast \end{pmatrix}  \mapsto U^\mathrm{HF} \begin{pmatrix} h_i \\ h_i^\ast \end{pmatrix} \equiv \begin{pmatrix}
\mathcal{U} & 0 \\
0 & \mathcal{U}^\ast
\end{pmatrix} \begin{pmatrix} h_i \\ h_i^\ast \end{pmatrix},
\end{equation}
\item or a GCP transformation~\cite{Lee:1966ik,Ecker:1981wv,Ecker:1987qp,Neufeld:1987wa} (these interchange the fields with their conjugates), 
\begin{equation}\label{Eq:H_to_GCP}
\begin{pmatrix} h_i \\ h_i^\ast \end{pmatrix} \mapsto U^\mathrm{GCP} \begin{pmatrix} h_i \\ h_i^\ast \end{pmatrix} \equiv \begin{pmatrix}
0 & \mathcal{U} \\
\mathcal{U}^\ast & 0
\end{pmatrix} \begin{pmatrix} h_i \\ h_i^\ast \end{pmatrix}.
\end{equation}
\end{itemize}

The group $U(3)$ contains the normal subgroup $U(1)$ of common phase rotations, which acts trivially on the scalar potential. Using $\cong$ to denote a group isomorphism, quotienting by this subgroup yields $U(3)/U(1)\cong PSU(3)$. Equivalently, one may quotient $SU(3)$ by its center $Z(SU(3))\cong\mathbb{Z}_3$, which also acts trivially, yielding the same projective group. Consequently, the search for symmetry groups of the 3HDM reduces to the study of finite subgroups of $PSU(3)$. More generally,
\begin{equation*}
U(N)\cong \left(SU(N)\times U(1) \right)/ \mathbb{Z}_N, \quad\text{so that}\quad U(N)/U(1)\cong SU(N)/\mathbb{Z}_N = PSU(N).
\end{equation*}

The HF and GCP transformations can be compactified into a single operator acting on the extended space:
\begin{equation}\label{Eq:W_normal}
\mathcal W(\mathcal{U}, \eta) = 
 \begin{pmatrix} \mathcal{U} & 0 \\
  0 & \mathcal{U}^* \end{pmatrix}\begin{pmatrix}
  (1-\eta)\, \mathcal{I} & \eta\, \mathcal{I} \\
   \eta\, \mathcal{I} & (1-\eta)\, \mathcal{I} \end{pmatrix},\text{ where } \eta=\{0,\,1\}.
\end{equation}
The choice of $\eta=0$ corresponds to an HF transformation of eq.~\eqref{Eq:H_to_HF}, while $\eta=1$ yields a GCP transformation of eq.~\eqref{Eq:H_to_GCP}. The matrix composed of $\eta$ generates $\mathbb{Z}_2$. Also, $\mathcal{I}$ is an identity matrix; since its dimensionality is usually clear from the context, we shall drop explicit identification of its dimensionality.

Building on the comprehensive classification of $U(3)$ non-isomorphic finite subgroups (up to order 2000) established in Ref.~\cite{Jurciukonis:2017mjp}, a recent re-evaluation of 3HDMs~\cite{Kuncinas:2025uty} has systematically mapped all known realisable symmetry groups. By utilising the $\mathsf{SmallGrp}$ library \cite{SmallGrp} within $\mathsf{GAP}$~\cite{GAP4}, the study classified these groups through their generators and irreducible representations. This analysis effectively concludes the exploration of conventional HF and GCP symmetry groups in 3HDMs:
\begin{itemize}
\item Realisable HF symmetry transformations:
\begin{align*}
U(1)_2, \quad \Delta(54)/ \mathbb{Z}_3, \quad S_3, \quad \mathbb{Z}_3, \quad \mathbb{Z}_2 \times \mathbb{Z}_2, \quad \mathbb{Z}_2.
\end{align*}

\item A class of (non-)realisable HF symmetry transformations that is also invariant under the trivial CP transformation (some complex couplings can be shown to be redundant):
\begin{align*}
& SU(3) \rtimes \mathbb{Z}_2^\ast, \quad SO(3) \times \mathbb{Z}_2^\ast, \quad \left[ \left[ U(1) \times U(1) \right] \rtimes S_3 \right] \rtimes \mathbb{Z}_2^\ast, \quad U(2) \rtimes \mathbb{Z}_2^\ast,\\
& \left[ O(2) \times U(1) \right]\rtimes \mathbb{Z}_2^\ast, \quad \left[ \left[ U(1) \times D_4 \right]/ \mathbb{Z}_2 \right] \rtimes \mathbb{Z}_2^\ast~\footnotemark,  \quad O(2) \times \mathbb{Z}_2^\ast\\
&  \left[ U(1) \times U(1) \right] \rtimes \mathbb{Z}_2^\ast, \quad \left[ U(1) \times \mathbb{Z}_2 \right] \rtimes \mathbb{Z}_2^\ast, \quad U(1)_1 \rtimes \mathbb{Z}_2^\ast \\
& \Sigma(36) \rtimes \mathbb{Z}_2^\ast, \quad S_4 \times \mathbb{Z}_2^\ast, \quad A_4 \rtimes \mathbb{Z}_2^\ast, \quad D_4 \times \mathbb{Z}_2^\ast, \quad \mathbb{Z}_4 \rtimes \mathbb{Z}_2^\ast.
\end{align*}
\footnotetext{Technically, assuming HF transformations, this symmetry group is described by $\widetilde{\mathcal{G}} \cong U(1) \circ_{Z(\mathcal{G})} \mathcal{G}$ with $Z(\mathcal{G}) \cong \mathbb{Z}_n, ~ \mathcal{G}/Z(\mathcal{G}) \cong V_4, ~ n \in 2\mathbb{Z}$; projectively it is given by $\left( U(1) \times D_4 \right) / Z(D_4) \cong U(1) \times V_4$. In Ref.~\cite{Kuncinas:2025uty} it was denoted as  $U(1) \circ V_4$.}

\item Extensions of realisable HF transformations by a trivial CP transformation:
\begin{align*}
& \quad U(1)_2 \rtimes \mathbb{Z}_2^\ast, \quad \left[ \Delta(54)/ \mathbb{Z}_3 \right] \rtimes \mathbb{Z}_2^\ast, \quad S_3 \rtimes \mathbb{Z}_2^\ast , \quad \mathbb{Z}_3 \rtimes \mathbb{Z}_2^\ast,\\
& \quad \mathbb{Z}_2 \times \mathbb{Z}_2 \times \mathbb{Z}_2^\ast, \quad \left[ \mathbb{Z}_2 \times \mathbb{Z}_2 \right] \rtimes \mathbb{Z}_2^\ast, \quad \mathbb{Z}_2 \rtimes \mathbb{Z}_2^\ast.
\end{align*}

\item Realisable GCP symmetry transformations:
\begin{align*}
\mathbb{Z}_2^\ast, \quad \mathbb{Z}_4^\ast.
\end{align*}

\end{itemize}
Any further imposition of additional groups reverts to established structures.

Let us briefly address a conceptual question: what happens to the scalar potential if we strictly decouple the bilinear sector, $V_2 =0$, leaving only $V_4$? The symmetry of the scalar potential is defined by the intersection of the symmetries of its quadratic and quartic components: $G_{\text{full}} = G_{V_2} \cap G_{V_4}$. These individual symmetry groups do not always coincide. For instance, imposing either a $\mathbb{Z}_2$ or a $U(1)$ symmetry can, depending on the choice of generators, lead to the same diagonal form of the quadratic potential, while generating entirely different sets of invariant terms in the quartic sector. It is natural to ask whether the converse is true: could two different symmetry groups give rise to the same quartic potential $V_4$, while inducing structurally different quadratic terms in $V_2$? Under the standard symmetry assumptions, this is strictly impossible.

This occurs because the trace of the bilinear matrix, $\mathrm{tr}(\mathcal{H}) = \sum_i h_{ii}$, is a universal singlet under basis transformations. Whenever a symmetry allows a specific invariant structure in $V_2$, it inevitably couples with the trace to produce a corresponding invariant configuration in $V_4$.

The only mechanism to bypass this constraint, thereby allowing different symmetry groups to yield the exact same $V_4$ whilst rendering different $V_2$ sectors, is if the transformations act non-trivially on the field trace itself. If the applied symmetries induce a transformation where $\mathrm{tr}(\mathcal{H})$ is not an invariant singlet, this necessitates that cross-terms in $V_4$ be strictly forbidden, allowing the quartic sector to decouple from the constraints of the quadratic parameter space.

Consider an operation where the scalar fields are $\mathbb{Z}_2$-even singlets, whilst the conjugated fields transform as $\mathbb{Z}_2$-odd quantities. This operation maps every fundamental $SU(2)$ bilinear scalar product to its negative: $h_{ij} \mapsto - h_{ij}$. The quartic sector is entirely ``blind" to this sign flip, since $h_{ij}$ undergo a double negative inversion. On the other hand, the quadratic terms $\mu_{ij}^2 h_{ij}$ contain only a single field contraction and thus transform as odd quantities. Therefore, invariance requires that $V_2=0$ while $V_4$ is completely general.

From a model-building perspective, vanishing of the quadratic couplings is not an abstract limit, but finds a realisation in the Scale Invariant (SI) theory~\cite{Bardeen:1995kv}. The central idea is to forbid explicit mass terms in the Lagrangian, thereby rendering the theory classically SI at tree level. This is then broken radiatively, allowing electroweak symmetry breaking to arise dynamically through the Coleman-Weinberg mechanism~\cite{Coleman:1973jx}, which was extended to multi-scalar cases by Gildener and Weinberg~\cite{Gildener:1976ih}. Consequently, the special GOOFy limit where all quadratic couplings are forced to vanish coincide with the Gildener-Weinberg class of models. For further reading on the implementation of SI models within the NHDM literature, particularly within the 2HDM framework, see Refs.~\cite{Lee:2012jn, Hashino:2015nxa, Lane:2019dbc, Braathen:2020vwo, Eichten:2022vys, Slavich:2026zmi}.

\section{Behind the GOOFy transformations}\label{Sec:GOOFY_transformations}

Before formalising the extended parameter space of GOOFy transformations, we briefly review their identification. Crucially, unlike conventional symmetries, this enlarged algebra permits scalar fields and their conjugates to transform under different representations. Whether viewed as physical symmetries or purely algebraic tools, they require a mathematically consistent, unified framework with a well-defined action on invariant operators.

\subsection{The GOOFy framework}

The defining feature of a GOOFy transformation is that it explicitly violates the usual kinetic invariance. Consequently, the field transformations alter the kinetic structure of the Lagrangian, a property that demands a fundamental reassessment of how symmetries are defined and enforced within multi-scalar sectors.

Originally, Ref.~\cite{Ferreira:2023dke} identified within the 2HDM a set of RG-stable parameter relations:
\begin{equation}\label{Eq:r0_2hdm_Couplings}
m_{11}^2 + m_{22}^2 = 0, \qquad \lambda_1 -\lambda_2 = 0, \qquad \lambda_6 + \lambda_7 = 0,
\end{equation}
written in the standard 2HDM notation and structurally reminiscent of the CP2-invariant model. The associated field transformation,
\begin{equation}\label{Eq:r0_2HDM}
h_{ij} \mapsto \begin{pmatrix}
-h_{22} & h_{21} \\
h_{12} & -h_{11}
\end{pmatrix}, \qquad
\begin{aligned}
h_1 &\mapsto -h_2^*, \qquad & h_1^\dagger &\mapsto h_2^{\mathrm T}, \\
h_2 &\mapsto h_1^*, \qquad  & h_2^\dagger &\mapsto -h_1^{\mathrm T},
\end{aligned}
\end{equation}
was shown to correspond to a purely imaginary scaling of the fields~\cite{Ferreira:2025ate}. 

The origin of these parameter constraints becomes transparent in the bilinear formalism~\cite{Velhinho:1994np,Nagel:2004sw,Ivanov:2005hg,Maniatis:2006fs,Ivanov:2006yq,Maniatis:2006jd,Maniatis:2007vn,Ivanov:2007de,Nishi:2007dv}. Gauge-invariant contractions of the doublets are encoded by
\begin{equation}
r_\mu = \frac{1}{2} h^\dagger \sigma_\mu h, \qquad \sigma_\mu = \left(\mathcal{I},\, \sigma_1,\, \sigma_2,\, \sigma_3\right),
\end{equation}
so that the scalar potential can be written as
\begin{equation}
V = M_0 r_0 + M_i r_i + \Lambda_{00} r_0^2 + \Lambda_{0i} r_0 r_i + \Lambda_{ij} r_i r_j,
\end{equation}

The $r_0$ component captures the overall positive-semidefinite norm of the scalar field configuration. By contrast, the $\vec r$ components encode the symmetry-sensitive directions of field space and may vanish for particular symmetry realisations. Conventional HF transformations in the 2HDM act as $SO(3)$ rotations on $\vec r$, while GCP transformations span the $O(3)$ group. In both cases, the $r_0$ direction remains invariant.

By contrast, consider the formal transformation
\begin{equation}
r_0 \to -r_0.
\end{equation}
Requiring invariance of the scalar potential under this transformation eliminates all terms odd in $r_0$, implying
$M_0 = 0$ and $\Lambda_{0i} = 0$. In the 2HDM, these constraints coincide with the defining relations of the $r_0$-symmetric model introduced in Ref.~\cite{Ferreira:2023dke}, namely eq.~\eqref{Eq:r0_2hdm_Couplings}.

The scalar potential relations were found to be stable under quantum corrections. In Ref.~\cite{Ferreira:2023dke} it was demonstrated that conditions
\begin{equation}
\beta_{(m_{11}^2+m_{22}^2)}=0, \qquad \beta_{(\lambda_1-\lambda_2)}=0, \qquad \beta_{(\lambda_6+\lambda_7)}=0,
\end{equation}
define an RG-stable manifold, at least through the three-loop scalar beta functions~\cite{Bednyakov:2018cmx} (see also Ref.~\cite{Bednyakov:2025sri} for an extended mathematical discussion involving basis-invariant syzygies). However, at the one-loop level, different analyses in the literature reach different conclusions regarding its validity.

In Ref.~\cite{Ferreira:2025ate}, the $r_0$-symmetry is interpreted as an extended imaginary scaling acting simultaneously on the fields, spacetime coordinates, and the renormalisation scale itself, $\Lambda_{\rm UV}^2 \mapsto -\Lambda_{\rm UV}^2$. By forcing the unphysical regularisation scale to transform alongside the physical variables, the one-loop effective potential preserves the RG relations. In this framework, $r_0$ behaves not as an internal symmetry, but as a generalised spacetime scaling covariance. By contrast, Ref.~\cite{Pilaftsis:2024uub} strictly analyses the effective potential within a standard physical framework where matching scales are held constant. In the dimensional regularisation scheme, underlying quadratic divergences re-emerge as finite, physical threshold corrections. These finite terms introduce odd powers of the symmetry-breaking parameter, explicitly violating the $r_0$ invariance at the one-loop level. The differing conclusions therefore originate from distinct treatments of the renormalisation scale rather than from the tree-level structure of the scalar potential itself.

\subsection[Obstacles to extending the \texorpdfstring{$r_0$}{r0} framework]{Obstacles to extending the $r_0$ framework}

In the 3HDM, imposing conditions based solely on the $r_0 \to -r_0$ transformation ($M_0=0$\linebreak and $\Lambda_{0a}=0$) does not define an RG-invariant subspace: already at one loop, the corresponding $\beta$ functions receive non-vanishing contributions from the $r_0$-even sector, thereby regenerating $r_0$-odd parameters under the RG flow. This indicates that the RG stability of the $r_0 \to -r_0$ manifold in the 2HDM relies on a special property of $SU(2)$ rather than a general feature of the NHDMs.

The bilinear formalism for the 3HDM is summarised in Appendix~\ref{App:Bilinear_3HDM} and RG conditions will be discussed in Subsection~\ref{Sec:Evaluating_RG}. Throughout this section we compute the $\beta$ functions up to two loops (excluding fermions) using \textsf{PyR@TE}~\cite{Sartore:2020gou} and \textsf{RGBeta}~\cite{Thomsen:2021ncy}.

A simple illustration is obtained by naively lifting the 2HDM condition of eq.~\eqref{Eq:r0_2hdm_Couplings} to the 3HDM. One finds
\begin{equation}
\begin{aligned}
\beta_{(\mu_{11}^2+\mu_{22}^2)} \propto{}& 2 \left( 3 \lambda_{1113} + \lambda_{1223} + 2 \lambda_{1322} \right) \mu_{13}^2 \\
&  + 2 \left( 2 \lambda_{1123} + \lambda_{1321} + 3 \lambda_{2223} \right) \mu_{23}^2 \\
&  + \left( 2 \lambda_{1133} + \lambda_{1331} + 2 \lambda_{2233} + \lambda_{2332} \right) \mu_{33}^2,
\end{aligned}
\end{equation}
demonstrating that the lifted condition is violated by radiative corrections. While the absence of $\mu_{12}^2$ mirrors the isolated 2HDM structure, the remaining contributions arise from interactions involving the $h_3$ doublet.

More generally, the conditions corresponding to $M_0=0$ and $\Lambda_{0a}=0$ are:
\begin{subequations}
\begin{align}
\mu_{11}^2 + \mu_{22}^2 + \mu_{33}^2 &= 0,\\
\lambda_{1112} + \lambda_{1222} + \lambda_{1233} &= 0,  \\
\lambda_{1113} + \lambda_{1322} + \lambda_{1333} &= 0,  \\
\lambda_{1123} + \lambda_{2223} + \lambda_{2333} &= 0,  \\
2\lambda_{1111} - 2\lambda_{2222} + \lambda_{1133} - \lambda_{2233} &= 0, \\
2\lambda_{2222} - 2\lambda_{3333} + \lambda_{1122} - \lambda_{1133} &= 0.
\end{align}
\end{subequations}
However, these relations likewise fail to define an RG-invariant manifold, as illustrated by
\begin{equation}
\begin{aligned}
\beta_{(\mu_{11}^2+\mu_{22}^2+\mu_{33}^2)} \propto{}& \big( 2\lambda_{1111} - \lambda_{1122} + \lambda_{1133} + \lambda_{1221} - 2\lambda_{2222} - \lambda_{2332} \big) \mu_{11}^2  \\
&+ 2 \big( \lambda_{1112} + \lambda_{1222} + \lambda_{1332} \big) \mu_{12}^2  \\
&+ 2 \big( \lambda_{1223} - \lambda_{1322} \big) \mu_{13}^2  \\
&+ \big( -\lambda_{1122} + \lambda_{1133} + \lambda_{1221} - \lambda_{1331} \big) \mu_{22}^2  \\
&+ 2 \big( -\lambda_{1123} + \lambda_{1321} \big) \mu_{23}^2.
\end{aligned}
\end{equation}

In the bilinear formalism the vector $r_a$ transforms in the adjoint representation of $SU(N)$, while the quartic couplings form a symmetric adjoint tensor, $\Lambda_{ab} \in \mathrm{Sym}^2(\mathrm{Adj})$. Consequently, the RG equations may contain every invariant tensor contraction permitted by the $SU(N)$ algebra. The crucial distinction between the 2HDM and higher NHDMs is the existence of the symmetric invariant tensor $d_{abc}$, defined through
\begin{equation}
\{t_a,t_b\} = \frac{1}{N}\delta_{ab}\mathcal{I} + d_{abc}\,t_c,
\end{equation}
where $t_a$ are the generators of $SU(N)$. Unlike the antisymmetric structure constants $f_{abc}$, the symmetric tensor $d_{abc}$ defines invariant symmetric contractions of adjoint tensors that can contribute to the RG flow.

For $SU(2)$, the adjoint representation is $\mathbf{3}$. The tensor product
\begin{equation}
\mathbf{3}\otimes\mathbf{3} = \left(\mathbf{1}\oplus\mathbf{5}\right)_S \oplus \left(\mathbf{3}\right)_A
\end{equation}
contains no adjoint component within its symmetric subspace, equivalently $d_{abc}=0$. Consequently, invariant tensor structures requiring a symmetric map $\mathrm{Sym}^2(\mathrm{Adj})\rightarrow\mathrm{Adj}$ are absent in $SU(2)$ because $\mathrm{Adj}\not\subset\mathrm{Sym}^2(\mathrm{Adj})$. In contrast, such structures are present in $SU(N\ge3)$, where $\mathrm{Adj}\subset\mathrm{Sym}^2(\mathrm{Adj})$.

By contrast, the adjoint representation of $SU(3)$ obeys
\begin{equation}
\mathbf{8}\otimes\mathbf{8} = \left(\mathbf{1} \oplus \mathbf{8} \oplus\mathbf{27} \right)_S \oplus \left( \mathbf{8}\oplus \mathbf{10} \oplus \overline{\mathbf{10}} \right)_A,
\end{equation}
and therefore contains the adjoint representation within its symmetric subspace, implying $d_{abc}\neq0$. The presence of $d_{abc}$ permits additional invariant contractions between quadratic and quartic couplings, thereby introducing tensor structures that contribute to the RG equations and are absent in the 2HDM. This explains why the naive lift of the $r_0$ conditions fails to remain invariant under the RG flow.

Because the existence of such a symmetric map is equivalent to the existence of the invariant tensor $d_{abc}$, it suffices to examine the invariant contraction
\begin{equation}
d_{abc}d^{abc}=\frac{(N^2-1)(N^2-4)}{N},
\end{equation}
which vanishes for $N=2$. Thus, every $SU(N\ge3)$ algebra possesses this non-vanishing symmetric invariant tensor. While this observation alone does not classify all RG-invariant submanifolds, it demonstrates that the isolated $r_0$ stability mechanism of the 2HDM relies on an exceptional algebraic property of $SU(2)$ rather than being a generic feature of NHDMs.

As argued in Refs.~\cite{Pilaftsis:2024uub,Pilaftsis:2026wyw}, the 2HDM $r_0$ construction should be interpreted not as a conventional field-space symmetry, but rather as an algebraic parity (or involution) acting on parameter space. This perspective is supported by recent results in simpler two-scalar models, where RG-stable GOOFy relations can be reproduced by imposing ordinary symmetries within an extended UV theory with twice the field content~\cite{Haber:2025cbb}. Such examples demonstrate that non-standard parameter relations may arise as effective low-energy remnants of a larger UV structure. In this light, our results admit two distinct physical interpretations. If the observed scalar sector is merely the low-energy limit of a $SU(N\ge3)$ UV theory, the isolated $r_0$ relations may be understood as an effective consequence of truncation to the $SU(2)$ algebra, in which the symmetric tensor $d_{abc}$ vanishes. If, instead, the 2HDM constitutes the fundamental scalar sector, then this exceptional algebraic protection may point to a special organising principle intrinsic to the theory.

The above observations connect with several recent approaches to the study of RG-invariant manifolds. In the outer-automorphism formalism~\cite{deBoer:2026ktb} (see also Ref.~\cite{Doring:2024kdg} for the discrete realisable symmetry groups of the 3HDM), the existence of a consistent outer automorphism of the global symmetry structure provides a sufficient criterion for the construction of RG-invariant parameter manifolds. A complementary perspective is offered by the spurion formalism~\cite{Pilaftsis:2026wyw}, in which symmetry-breaking parameters are promoted to spurion fields and the stability of a parameter manifold is determined by the existence of allowed covariant spurion contractions. Consequently, an isolated $r_0$-type parameter parity does not appear to be sufficient to ensure RG stability, and additional symmetry constraints are generally required to eliminate the corresponding invariant structures~\cite{deBoer:2026ktb,Pilaftsis:2026wyw}. While such constraints can stabilise particular parameter submanifolds, they should not be regarded as a general classification of RG-invariant manifolds in arbitrary NHDMs.

These observations motivate the use of the Hilbert series as a framework for classifying RG-invariant manifolds. The Hilbert series has been widely used across particle physics~\cite{Benvenuti:2006qr,Feng:2007ur,Jenkins:2009dy,Lehman:2015via,Henning:2017fpj}, including applications to the 2HDM~\cite{Trautner:2018ipq,Bednyakov:2025sri} and 3HDM~\cite{Bento:2021hyo,Bryan:2026xvs}. As a generating function, it enumerates the independent polynomial basis invariants associated with a given symmetry group, together with the algebraic relations (syzygies) among them. Since RG evolution is basis covariant, any RG-invariant manifold should be expressible in terms of these fundamental invariants.

\subsection{Defining GOOFy transformations}

To handle GOOFy transformations it is necessary to adopt an extended reducible space for the set of $N$ scalar doublets and their conjugates. We define $h$ as the $N$-dimensional column vector of the $SU(2)$ scalar doublets:
\begin{equation}
h = (h_1^\mathrm{T}, h_2^\mathrm{T}, \dots, h_N^\mathrm{T})^\mathrm{T}.
\end{equation}
Next, we construct an extended $2N$-dimensional vector space
\begin{equation}
H \equiv h \oplus h^\ast = \begin{pmatrix}
h \\
h^\ast
\end{pmatrix},
\end{equation}
alternatively, it can be expressed in a compact notation as $H = (h^\mathrm{T}, h^\dagger)^\mathrm{T}$. Note that $H^\mathrm{T} = (h^\mathrm{T}, h^\dagger)$.

A GOOFy transformation breaks the strict $\mathcal{U}-\mathcal{U}^*$ correspondence that constrains standard HF and GCP transformations. Specifically, these operations manifest as either:
\begin{itemize}
\item an HF-like transformation:
\begin{equation}\label{Eq:HF_GOOFy_tr}
H \mapsto \begin{pmatrix}
\mathcal{U}^{(1)} & 0 \\
0 & \mathcal{U}^{(2)}
\end{pmatrix} H,
\end{equation}
\item a GCP-like transformation:
\begin{equation}\label{Eq:GCP_GOOFy_tr}
H \mapsto \begin{pmatrix}
0 & \mathcal{U}^{(1)}\\
\mathcal{U}^{(2)} & 0
\end{pmatrix} H,
\end{equation}
\end{itemize}
subject to the condition $\mathcal{U}^{(2)} \neq (\mathcal{U}^{(1)})^*$. This independent action on the fields and their conjugates distinguishes GOOFy transformations from conventional HF and GCP symmetries and underlies their non-trivial properties.

\subsection{The kinetic Lagrangian}

The strict definition of a symmetry requires that the transformation leaves the entire Lagrangian invariant. For standard HF and GCP transformations, the kinetic term \linebreak $\mathcal{L}_{K} = (D_\mu h)^\dagger (D^\mu h)$ remains invariant because transformations amount to a unitary rotation in the scalar space. To understand how the kinetic term transforms under the GOOFy transformation, we define the extended gauge-kinetic metric $K$ as:
\begin{equation}
K = \begin{pmatrix} 0 & \mathcal{K}^* \\ \mathcal{K} & 0 \end{pmatrix}, \quad \text{where} \quad \mathcal{K} \equiv \overleftarrow{D_\mu^{\dagger\phantom{|}}} \overrightarrow{D^{\mu\phantom{|}}}.
\end{equation}

The kinetic Lagrangian can then be compactly written in the extended $H$ space:
\begin{equation}
\mathcal{L}_{K} = \frac{1}{2} H^\mathrm{T} K H.
\end{equation}

When applying a general $2N \times 2N$ transformation $H \mapsto \mathcal{W}_{12} H$, with 
\begin{equation}\label{Eq:W_GOOFy_U1_U2}
\mathcal{W}_{12}(\mathcal{U}^{(1)}, \mathcal{U}^{(2)}, \eta) = \begin{pmatrix}
\mathcal{U}^{(1)} & 0 \\
0 & \mathcal{U}^{(2)}
\end{pmatrix}\begin{pmatrix}
(1-\eta)\, \mathcal{I} & \eta\, \mathcal{I}  \\
\eta\, \mathcal{I} & (1-\eta)\, \mathcal{I}
\end{pmatrix},
\end{equation}
the kinetic Lagrangian transforms as
\begin{equation}
\mathcal{L}_{K} \mapsto \frac{1}{2} H^\mathrm{T} (\mathcal{W}_{12}^\mathrm{T} \,K \, \mathcal{W}_{12}) H.
\end{equation}

By expanding the product and isolating the lower-left block that pairs $h^\dagger \dots h$, we find the physical kinetic transformation:
\begin{equation}\label{Eq:L_K_phys_tr}
\mathcal{L}_{K,\,phys}^\prime = h^\dagger \Big[ (1-\eta) \mathcal{K} \mathcal{X} + \eta \mathcal{K}^\ast \mathcal{X}^\mathrm{T} \Big] h,
\end{equation}
where we have defined the anomaly matrix $\mathcal{X}$ as:
\begin{equation}\label{Eq:Def_X_GOOFy}
 \mathcal{X} = (\mathcal{U}^{(2)})^\mathrm{T}\, \mathcal{U}^{(1)}.
\end{equation}
Because $\mathcal{U}^{(1)}$ and $\mathcal{U}^{(2)}$ are unitary, their product $\mathcal{X}$ is also a unitary matrix.

By dropping the redundant superscript and defining the rotation as $\mathcal{U} \equiv \mathcal{U}^{(1)}$, the conjugate block is determined by $\mathcal{U}^{(2)} = \mathcal{U}^* \mathcal{X}^\mathrm{T}$. We can thus rewrite the GOOFy operator of eq.~\eqref{Eq:W_GOOFy_U1_U2} in terms of $\mathcal{U}$ and $\mathcal{X}$:
\begin{equation}\label{Eq:Def_W_GOOFy_U_X}
\mathcal{W}\,(\mathcal U,\, \mathcal{X},\, \eta) = \begin{pmatrix}
\mathcal{U} & 0 \\
0 & \mathcal{U}^\ast \end{pmatrix} \begin{pmatrix}
(1-\eta)\, \mathcal{I} & \eta\, \mathcal{I} \\
\eta\, \mathcal{X}^\mathrm{T} & (1-\eta)\, \mathcal{X}^\mathrm{T}
\end{pmatrix}.
\end{equation}
This structure allows us to separate the standard flavour rotation from the kinetic anomaly. In this form, one observes that the anomalous kinetic scaling can be written as a product of the standard HF transformation and the mixing operator $\mathcal{X}$. Without loss of generality, we adopt a convention where the primary symmetry action is assigned to the upper block; this effectively restricts the GOOFy-specific deviations to the conjugate sector.

Because a matrix and its transpose share identical eigenvalues, we can characterise the action on the kinetic sector through the eigenvalues $\chi_i$ of $\mathcal X$,
\begin{equation}\label{LK_under_GOOFy}
(D_\mu h_i)^\dagger (D^\mu h_i) \mapsto  \chi_i\, (D_\mu h_i)^\dagger (D^\mu h_i).
\end{equation}
For an exact symmetry of the kinetic Lagrangian one must require $\mathcal X=\mathcal I$. Relaxing this condition leads to several possibilities:
\begin{itemize}
\item Non-Hermiticity ($\chi_i \notin \mathbb{R}$): complex eigenvalues render a non-Hermitian kinetic Lagrangian, signalling a loss of unitarity;

\item  Kinetic signature with $\chi_i=-1$: a negative eigenvalue reverses the sign of the corresponding kinetic term. Within a conventional field-theoretic interpretation, this describes a ghost degree of freedom. However, Ref.~\cite{Trautner:2025yxz} argues that such sign flips should be treated as part of an enlarged algebraic transformation acting on the parameter space rather than as physical symmetries of the low-energy kinetic sector. For global sign-flipping transformations, the associated parameter relations remain stable to all orders, since the explicit breaking from the gauge-kinetic terms is soft. By contrast GOOFy transformations with a relative sign flip give rise to parameter relations that are not all stable under RG running;

\item Vanishing kinetic fields: if one demands invariance under a transformation where $\chi_i = -1$, the identity $\mathcal{L}_{\text{kin}} = -\mathcal{L}_{\text{kin}}$ forces the affected kinetic terms to vanish. Because the kinetic sector has no free parameters to absorb this sign, the affected doublets become auxiliary fields, effectively reducing the active dimensionality of the NHDM.
\end{itemize}

Considering the above options, we must restrict $\mathcal{X}$ to be Hermitian to preserve physical viability. Since $\mathcal{X}$ is both unitary and Hermitian, it must also satisfy the involution property $\mathcal{X}^2 = \mathcal{I}$. This strictly restricts the admissible eigenvalues to $\chi_i \in \{-1, +1\}$.

\subsection{Basis transformations}

Assume a basis rotation of the doublets, $h \to V h$. Applying the extended rotation $\mathcal{V} = \mathrm{diag}(V,\, V^\ast)$ to the operator $\mathcal{W}$ yields:
\begin{equation}
\mathcal{W}\xrightarrow{\mathcal{V}}{} \mathcal{W}' = \begin{pmatrix} (1-\eta) V^\dagger \mathcal{U} V & \eta V^\dagger \mathcal{U} V^* \\ \eta V^\mathrm{T} \mathcal{U}^* \mathcal{X}^\mathrm{T} V & (1-\eta) V^\mathrm{T} \mathcal{U}^* \mathcal{X}^\mathrm{T} V^* \end{pmatrix}.
\end{equation}
Then, the anomaly matrix $\mathcal{X}$ transforms as:
\begin{itemize}
\item For HF-like symmetries ($\eta=0$):
\begin{equation}
\mathcal{X} \to \mathcal{X}^\prime = V^\dagger \mathcal{X} V.
\end{equation}
\item For GCP-like symmetries ($\eta=1$):
\begin{equation}
\mathcal{X} \to \mathcal{X}^\prime = V^\mathrm{T} \mathcal{X} V^\ast.
\end{equation}
\end{itemize}

It is always possible to pick a basis transformation $V$ such that $\mathcal{X}$ becomes diagonal. However, rotating into this basis simultaneously alters the standard generator $\mathcal{U}$. When $\mathcal{U}$ and $\mathcal{X}$ do not commute, $[\mathcal{U}, \mathcal{X}] \neq 0$, it indicates that these transformations cannot share a simultaneous eigenbasis. While a choice of $\mathcal{X}=\pm \mathcal{I}$ alters only the quadratic couplings, a different choice of $\mathcal{X}$ can in general lead to a different quartic part of the potential. 

Instead of analysing how the individual scalar fields rotate, we can construct a scalar potential using the gauge-invariant bilinears $r_a \propto h^\dagger t_a h$, where $t_a$ are the generators of $SU(N)$. Applying the extended generator, the bilinears transform as:
\begin{equation}
r_a \mapsto r_a^\prime \propto (h^\dagger)' t_a h' \sim  h^\dagger \Big[ (1-\eta) \mathcal{X} \mathcal{U}^\dagger t_a \mathcal{U} + \eta \, \mathcal{U}^\mathrm{T} t_a^\mathrm{T} \mathcal{U}^\ast \mathcal{X}^\mathrm{T} \Big] h.
\end{equation}

When applying a standard symmetry, $\mathcal{X} = \mathcal{I}$, the term $\mathcal{U}^\dagger t_a \mathcal{U}$ describes an $O(N^2-1)$ adjoint rotation among the bilinear singlets. However, when $\mathcal{X} \neq \mathcal{I}$, this rotation is distorted. For the $SU(N)$ geometry to remain unbroken, the matrix $\mathcal{X}$ would need to commute with every single $SU(N)$ generator, $[\mathcal{X}, t_a] = 0$. The only matrices capable of this are multiples of the identity, $\mathcal{X} \propto \mathcal{I}$. If we select $\mathcal{X} = -\mathcal{I}$, the matrices commute but a universal minus sign is introduced; this uniformly flips the bilinears. Conversely, if we choose a representation that is not proportional to the identity, we guarantee that $\mathcal{X}$ will fail to commute with a specific subset of the $t_a$ generators. The non-commutativity acts as a filter: it introduces relative sign flips among specific bilinears.

In terms of the $h_{ij}$ notation, a basis change of $V$ yields the following relations:
\begin{subequations}\label{Eq:}
\begin{align}
\text{HF-like: } & \quad h_{ij} \to \sum_{m,n}  (V^\mathrm{T} \mathcal{U}^\ast \mathcal{X}^\mathrm{T} V^\ast)_{im} (V^\dagger \mathcal{U} V)_{jn} h_{mn},\\
\text{GCP-like: } & \quad h_{ij} \to \sum_{m,n} (V^\mathrm{T} \mathcal{U}^\ast \mathcal{X}^\mathrm{T} V)_{im} (V^\dagger \mathcal{U} V^\ast)_{jn} h_{nm}.
\end{align}
\end{subequations}

Consider an HF-like transformation, we can define a diagonal $\hat{\mathcal{X}} \equiv V^\dagger \mathcal{X} V$ matrix and a rotated matrix $\widetilde{\mathcal{U}} \equiv V^\dagger \mathcal{U} V$, which, in general is not diagonal. Then, $V^\mathrm{T} \mathcal{U}^\ast \mathcal{X}^\mathrm{T} V^\ast = \widetilde{\mathcal{U}}^\ast \hat{\mathcal{X}}$. For a GCP-like transformation one defines $\hat{\mathcal{X}} \equiv V^\mathrm{T} \mathcal{X} V^\ast$ and $\widetilde{\mathcal{U}} \equiv V^\dagger \mathcal{U} V^\ast$. As a result, in the $\hat{\mathcal X}$ basis we have:
\begin{subequations}
\begin{align}
\text{HF-like: } & \quad h_{ij} \to \sum_{m,n} \hat{\mathcal{X}}_{mm} \widetilde{\mathcal{U}}_{im}^\ast \widetilde{\mathcal{U}}_{jn} h_{mn},\\
\text{GCP-like: } & \quad h_{ij} \to \sum_{m,n} \hat{\mathcal{X}}_{mm} \widetilde{\mathcal{U}}_{im}^\ast \widetilde{\mathcal{U}}_{jn} h_{nm}.
\end{align}
\end{subequations}

To sum up, if one fixes the $\mathcal{U}$ generator to investigate a specific symmetry, applying an extended  basis transformation $\mathcal V$ to force $\mathcal{X}$ into a diagonal frame will inevitably transform $\mathcal{U}$. Such basis transformation might obscure the manifest representation.

While the transformations of $h_{ij}$ derived above describe the effects of a basis change, they simultaneously define the conditions for a symmetry invariance condition. If the scalar potential is demanded to be invariant under these mappings, the rotated parameter matrices must be identical to the original ones. For a general $\mathcal{X}$, couplings transform as:
\begin{align}
\text{HF-like: } \quad \mu_{ij}^2 \to \mu_{mn}^2 ={}& \sum_{i,j,a} \mathcal{X}_{ma} \mathcal{U}_{ia}^\ast  \mathcal{U}_{jn} \mu_{ij}^2, \\
\text{GCP-like: } \quad \mu_{ij}^2 \to \mu_{mn}^2 ={}& \sum_{i,j,a} \mathcal{X}_{na} \mathcal{U}_{ja}^\ast  \mathcal{U}_{im} (\mu_{ij}^{2})^\ast.
\end{align}
Similarly, the quartic tensor $\lambda_{ijkl}$ is subject to
\begin{align}
\text{HF-like: } \quad \lambda_{ijkl} \to \lambda_{mnop} ={}& \sum_{i,j,k,l,a,b} \mathcal{X}_{ma} \mathcal{X}_{ob} \mathcal{U}_{ia}^\ast \mathcal{U}_{jn} \mathcal{U}_{kb}^\ast \mathcal{U}_{lp} \lambda_{ijkl}, \\
\text{GCP-like: } \quad \lambda_{ijkl} \to \lambda_{mnop} ={}& \sum_{i,j,k,l,a,b} \mathcal{X}_{na} \mathcal{X}_{pb} \mathcal{U}_{ja}^\ast \mathcal{U}_{im} \mathcal{U}_{lb}^\ast \mathcal{U}_{ko} \lambda_{ijkl}^\ast.
\end{align}

Note that while the quadratic $\mu_{ij}^2$ term transforms alongside a single insertion of $\mathcal{X}$, the quartic coupling $\lambda_{ijkl}$ transforms alongside two insertions.

For a general analysis, it is computationally convenient to fix a specific $\mathcal{U}$ matrix and systematically evaluate the impact of different $\hat{\mathcal{X}}$ configurations. This approach circumvents the need to explicitly check the eigenvalues of $\mathcal{X}$ at each step, as only a limited set of admissible $\hat{\mathcal{X}}$ matrices exist up to permutations. Consequently, to simplify the notation throughout the remainder of this work, we adopt the convention where $\mathcal{X}$ is chosen to be diagonal, allowing us to drop the hat notation entirely.

\section{Admissible GOOFy transformations}\label{Sec:GOOFY_admissible}

Having established the extended multiplet space and its kinetic anomalies, we now systematically map these transformations onto the scalar potential. We begin by categorising the distinct classes of GOOFy operations and identifying the complete space they span.

Before proceeding, it might be instructive to take a look at several examples provided in Appendix~\ref{App:V2_GOOFy_options}.

\subsection{Closure of the extended transformation algebra}\label{Eq:G_of_WGOOFy}

To systematically classify the extended GOOFy transformations, we first study their closure under composition. As discussed in the previous section, we restrict $\mathcal X$ to be diagonal, and consider the action of the extended operator $\mathcal W$ on the $H$ space.

For two successive transformations of the form given in eq.~\eqref{Eq:Def_W_GOOFy_U_X}, one finds:
\begin{equation}
\mathcal{W}_1 \mathcal{W}_2 = \begin{pmatrix}
(1-\eta_1)\mathcal{U}_1 & \eta_1\mathcal{U}_1 \\
\eta_1\mathcal{U}_1^\ast\mathcal{X}_1 & (1-\eta_1)\mathcal{U}_1^\ast\mathcal{X}_1
\end{pmatrix}
\begin{pmatrix}
(1-\eta_2)\mathcal{U}_2 & \eta_2\mathcal{U}_2 \\
\eta_2\mathcal{U}_2^\ast\mathcal{X}_2 & (1-\eta_2)\mathcal{U}_2^\ast\mathcal{X}_2
\end{pmatrix} = \begin{pmatrix} A & B \\ C & D \end{pmatrix},
\end{equation}
where the block components are given by:
\begin{subequations}
\begin{align}
A &= (1-\eta_1)(1-\eta_2) \mathcal{U}_1\mathcal{U}_2 + \eta_1\eta_2 \mathcal{X}_2\mathcal{U}_1\mathcal{U}_2^\ast, \\
B &= (1-\eta_1)\eta_2 \mathcal{U}_1\mathcal{U}_2 + \eta_1(1-\eta_2) \mathcal{X}_2\mathcal{U}_1\mathcal{U}_2^\ast, \\
C &= \eta_1(1-\eta_2) \mathcal{X}_1\mathcal{U}_1^\ast\mathcal{U}_2 + (1-\eta_1)\eta_2 \mathcal{X}_1 \mathcal{X}_2 \mathcal{U}_1^\ast\mathcal{U}_2^\ast, \\
D &= \eta_1\eta_2 \mathcal{X}_1 \mathcal{U}_1^\ast\mathcal{U}_2 + (1-\eta_1)(1-\eta_2) \mathcal{X}_1 \mathcal{X}_2 \mathcal{U}_1^\ast\mathcal{U}_2^\ast.
\end{align}
\end{subequations}

Two important special cases follow:
\begin{itemize}
\item For HF-like transformations, $\mathcal{W}_{\text{HF}}^k = \mathcal{U}^k \oplus (\mathcal{U}^\ast \mathcal{X})^k$;

\item For GCP-like transformations,  $\mathcal{W}_{\text{GCP}}^2 = \text{diag}(\mathcal{X}\,\mathcal{U} \mathcal{U}^\ast ,\, \mathcal{X}\, \mathcal{U}^\ast  \mathcal{U})$.
\end{itemize}

To isolate the discrete structure introduced by the $\mathcal X$ block, it is sufficient to set $\mathcal U=\mathcal I$. Restricting to $\mathcal X=\pm\mathcal I$, one obtains three non-trivial metric operators acting on the extended $H$ space:
\begin{subequations}
\begin{align}
    \mathcal{K}_O &\equiv \mathcal{W}\,(\mathcal{I}, \mathcal{I}, 1) = \begin{pmatrix} 0 & \mathcal{I} \\ \mathcal{I} & 0 \end{pmatrix}, \\
    \mathcal{K}_P &\equiv \mathcal{W}\,(\mathcal{I}, -\mathcal{I}, 0) = \begin{pmatrix} \mathcal{I} & 0 \\ 0 & -\mathcal{I} \end{pmatrix}, \\
    \mathcal{K}_S &\equiv \mathcal{W}\,(\mathcal{I}, -\mathcal{I}, 1) = \begin{pmatrix} 0 & \mathcal{I} \\ -\mathcal{I} & 0 \end{pmatrix}.
\end{align}
\end{subequations}
Note that these operators are not independent, $\mathcal{K}_S \mathcal{K}_O = \mathcal{K}_P$. Furthermore, $\mathcal K_O^2=\mathcal K_P^2=\mathcal I_6$ and $(\mathcal K_P\mathcal K_O)^2=-\mathcal I_6$. Consequently, the subgroup generated by $\mathcal K_O$ and $\mathcal K_P$ is isomorphic to the dihedral group of order eight,
\begin{equation}\label{Eq:GOOFy_D4_full}
    D_4 \cong \Big\langle \mathcal{K}_O, \mathcal{K}_P ~\Big|~ \mathcal{K}_O^2 = \mathcal{K}_P^2 = (\mathcal{K}_P \mathcal{K}_O)^4 = \mathcal{I}_6 \Big\rangle.
\end{equation}
Its center is $ Z(D_4)=\langle-\mathcal I_6\rangle\cong\mathbb Z_2$, and factoring out this center gives $ D_4/Z(D_4) \cong \mathbb Z_2\times\mathbb Z_2$.

The metric operators define three forms on the $H$ space. A transformation $U$ is said to preserve the corresponding metric if
\begin{equation}
U^T\mathcal K_{S,O}U
=
\pm\mathcal K_{S,O},
\qquad
U^\dagger\mathcal K_PU
=
\pm\mathcal K_P.
\end{equation}

When evaluating the four discrete classes of transformations generated by the combinations of $\eta$ and $\mathcal{X}=\pm\mathcal{I}$ against these three metrics, a $\mathbb{Z}_2 \times \mathbb{Z}_2$ structure emerges:

\begin{table}[H]
\centering
\renewcommand{\arraystretch}{1.5}
\begin{tabular}{|c|c|c|c|}
\hline
Transformation & Parity under $\mathcal{K}_S$ & Parity under $\mathcal{K}_O$ & Parity under $\mathcal{K}_P$ \\
\hline
$U^\mathrm{HF}$ & $+1$ & $+1$ & $+1$ \\
$U^\mathrm{GCP}$ & $-1$ & $+1$ & $-1$ \\
$U^\mathrm{HF}_\pm$ & $-1$ & $-1$ & $+1$ \\
$U^\mathrm{GCP}_\pm$ & $+1$ & $-1$ & $-1$ \\
\hline
\end{tabular}
\end{table}

The discrete sector is therefore generated by two independent involutions, with the third obtained from their product. This yields a $\mathbb Z_2\times\mathbb Z_2$ grading of the extended transformation algebra. Combining this discrete structure with the continuous HF transformations suggests the enlarged transformation group
\begin{equation}
\mathcal G_{\mathrm{max}} = PSU(3) \rtimes (\mathbb Z_2\times\mathbb Z_2),
\end{equation}
where the discrete factor acts on the doubled field space through the metric operators introduced above.

\subsection{The saturation of cyclic groups}

As established in Ref.~\cite{Ivanov:2011ae}, any finite Abelian group $\mathcal{G}$ realisable within an NHDM must satisfy $|\mathcal{G}| \le 2^{N-1}$. Specifically, for a cyclic group $\mathbb{Z}_p$ to be realisable, its order $p$ cannot exceed the $2^{N-1}$ threshold. Attempting to impose a group that violates this bound results in non-realisability, as the potential is forced to manifest a larger, accidental symmetry.

In the specific case of the 3HDM, the standard HF transformations saturate at $\mathbb{Z}_4$, while the largest realisable discrete symmetry group in 3HDM is $\Sigma(36) \cong (\mathbb{Z}_3 \times \mathbb{Z}_3) \rtimes \mathbb{Z}_4$ (more precisely $\Sigma(36)_{PSU(3)} \cong \Sigma(36\times 3)_{SU(3)} \big/ \mathbb{Z}_3$). For any $n \geq 5$, the imposition of a regular cyclic $\mathbb{Z}_n$ symmetry exhausts the available quartic invariants and the scalar potential becomes invariant under a continuous $U(1)$ group.

However, the disjointed block-structure of GOOFy transformations circumvents this limitation (considering transformations acting on the extended $6\times6$ $H$ space). Because the $\mathcal{W}$ generator decouples sub-blocks that operate under misaligned cyclic sub-groups, the transformations can achieve higher composite orders in the extended $6\times6$ space. Interestingly, while this circumvents the bounds, we shall identify in Section~\ref{Sec:Realisable_GOOFy} that the maximum realisable cyclic group produced by a single generator is still constrained to $\mathbb{Z}_4$ in the 3HDM.

\subsection{Representation mismatch and selection rules}\label{Sec:Reps_mismatch}

We have already noted that applying $\mathcal{W}\,(\mathcal{I}, -\mathcal{I}, 0):\,h_{ij} \mapsto -h_{ij}$ yields an SI 3HDM. While this case is phenomenologically interesting because it forces all quadratic terms to vanish ($\mu_{ij}^2 = 0$), the resulting tree-level potential is classically scale-invariant. Consequently, the stationary-point equations determine only the vacuum alignment in field space, leaving the absolute scale of symmetry breaking undetermined at tree level; this scale is subsequently generated at one loop via the Gildener--Weinberg mechanism. As will be discussed in Subsection~\ref{Sec:Overall_X_discussion}, such scenarios generically give rise to additional massless states.

It is therefore more instructive to consider cases with $V_2\neq 0$. In the standard setting, the fields $h_i$ transform in a unitary representation $R(g)$ of a group $\mathcal G$, while their conjugates transform in the conjugate representation $R(g)^\ast$. A GOOFy transformation modifies this relation by a one-dimensional character $\chi(g)$, so that
\begin{equation}\label{Eq:GOOFy_rep_th}
h \mapsto R(g)h, \qquad
h^\ast \mapsto \chi(g)\,R(g)^\ast h^\ast.
\end{equation}

For this transformation to remain consistent with the structure of $\mathcal G$, the factor $\chi(g)$ cannot be arbitrary. Specifically, $\chi(e)=1$ must hold, and composition of transformations implies
\begin{equation*}
\chi(g_1g_2)=\chi(g_1)\chi(g_2),\qquad \forall\, \{g_1,g_2\}\in\mathcal G.
\end{equation*}

Thus, $\chi:\mathcal G\to U(1)$ is a group homomorphism, \textit{i.e.}, a one-dimensional character of $\mathcal G$. In the present construction, Hermiticity restricts $\chi$ to the values of $\pm 1$.

The character $\chi$ has direct implications for the quadratic couplings, $Y = \mu_{ij}^2$:
\begin{equation}
h^\dagger Y h \mapsto \chi(g) \left(h^\dagger R(g)^\dagger Y R(g)h\right), \qquad \forall\, g\in\mathcal G.
\end{equation}
Note that $\chi(g)^{-1} = \chi (g)^\ast = \chi (g)$ are all equivalent in our context for $\mathcal{X}$.

Requiring invariance of \(V_2\) then gives
\begin{equation}\label{Eq:RYR_condition}
R(g)^\dagger Y R(g)=\chi(g)\,Y.
\end{equation}

The above condition is interpreted as a constraint on the representation space $R^\ast\otimes R$, \linebreak in which the quadratic matrix $Y$ resides. The $Y_{ij}$ coefficients are fixed parameters. Therefore it is required that $Y$ should transform in the one-dimensional representation $\chi$ of $\mathcal G$.

When $\chi$ is trivial, this reduces to the familiar conditions, and invariance of $Y$ is dictated by $\mathcal G$. For a non-trivial character, only those components of $R^\ast\otimes R$ transforming according to $\chi$ may appear in the quadratic potential.

Let us consider several examples of the realisable 3HDMs:
\begin{itemize}
\item The $A_4$ symmetry group has four irreducible representations, $\{\mathbf{1},\,\mathbf{1}',\,\mathbf{1}'',\,\mathbf{3}\}$, where $\mathbf{1}'$ and $\mathbf{1}''$ are complex one-dimensional representations. For the triplet $\mathbf{3}$ (which is the relevant assignment for a realisable $A_4$ symmetry in the 3HDM), the tensor product decomposes as $\mathbf{3}\otimes\mathbf{3} = \mathbf{1}\oplus\mathbf{1}'\oplus\mathbf{1}''\oplus\mathbf{3}_S\oplus\mathbf{3}_A$. Since $A_4$ contains only even permutations, it does not admit a non-trivial $\mathbb{Z}_2$-valued character. Consequently, eq.~\eqref{Eq:RYR_condition} can only be satisfied for $V_2=0$.

\item The $S_4$ symmetry group has five irreducible representations, $\{\mathbf{1},\,\mathbf{1}',\,\mathbf{2},\,\mathbf{3}_1,\,\mathbf{3}_2\}$, where $\mathbf{1}'$ denotes the sign representation. The tensor products $\mathbf{3}_1\otimes\mathbf{3}_1$ and  $\mathbf{3}_2\otimes\mathbf{3}_2$ decompose as $\mathbf{1}\oplus\mathbf{2}\oplus\mathbf{3}_1\oplus\mathbf{3}_2$, so that $\mathbf{1}'$ does not appear. Moreover, $\mathbf{3}_2\cong\mathbf{3}_1\otimes\mathbf{1}'$, while the mixed product $\mathbf{3}_1\otimes\mathbf{3}_2$ contains no trivial singlet. Hence no component of $R^\ast\otimes R$ transforms according to the sign character, and the GOOFy selection rule again enforces $V_2=0$.

\item The symmetry groups $S_3$ and $D_4$ do allow for a non-vanishing quadratic potential, since their reducible three-dimensional representations of the form $\mathbf{1}\oplus\mathbf{2}$ produce tensor products containing the required one-dimensional character. For $S_3$, the tensor decomposition is given by $\mathbf{2}\otimes\mathbf{2} = \mathbf{1}\oplus\mathbf{1}'\oplus\mathbf{2}$, which contains the sign representation. Consequently, a non-zero $Y$ may satisfy eq.~\eqref{Eq:RYR_condition}. Similarly, for $D_4$, the tensor decomposition is $\mathbf{2}\otimes\mathbf{2} = \mathbf{1}_{++}\oplus\mathbf{1}_{+-}\oplus\mathbf{1}_{-+}\oplus\mathbf{1}_{--}$, which contains several one-dimensional representations. It is therefore sufficient that one of them coincides with the character $\chi$.

\item The $Z_n$ symmetry group contains  $n$ one-dimensional irreducible representations. A non-trivial sign character $\chi:\mathbb{Z}_n\rightarrow\{\pm1\}$ exists if and only if there is a homomorphism onto $\mathbb{Z}_2$. Equivalently, $\mathbb{Z}_n$ must contain an element of order two, namely $g^{n/2}$, which is possible only for even $n$. Therefore, odd cyclic groups cannot support GOOFy quadratic terms.

\item The $U(1)$ group contains the element $e^{i\pi}=-1$, but this does not define a non-trivial $\mathbb{Z}_2$-valued character of the group. Consequently, the quadratic couplings are forced to vanish. This conclusion extends immediately to continuous groups whose relevant representations are complex. By contrast, continuous groups with real representations may still admit non-trivial characters. For example, $O(2)$ possesses the determinant homomorphism $\det:O(2)\rightarrow\{\pm1\}$, which provides the required sign character. On the other hand, although $O(3)$ (or, more precisely, $SO(3)$ \linebreak after quotienting by the hypercharge transformation) also admits the determinant $\mathbb{Z}_2$ structure, the tensor product $\mathbf{3}\otimes\mathbf{3} = \mathbf{1}\oplus\mathbf{3}\oplus\mathbf{5}$ does not contain an additional one-dimensional component transforming according to the required character. Consequently, among the continuous symmetry groups realised in the 3HDM, only those containing an $O(2)$ factor, such as $O(2)$ and $O(2)\times U(1)$, can accommodate a non-vanishing quadratic potential.

\end{itemize}

The existence of a non-vanishing bilinear sector $V_2$ compatible with a GOOFy transformation is governed by two independent conditions:
\begin{itemize}
\item There must exist a non-trivial homomorphism $\chi: \mathcal{G} \to \{\pm 1\}$. Equivalently, the Abelianisation $\mathcal{G}/[\mathcal{G},\mathcal{G}]$ must admit a non-trivial $\mathbb{Z}_2$ quotient;

\item The chosen scalar representation $R: \mathcal{G} \to U(3)$ must be such that the $\chi$ character appears in the decomposition of the tensor product $R^\ast \otimes R$. Equivalently, there must exist a non-zero matrix $Y$ satisfying $ R(g)^\dagger Y R(g) = \chi(g)\, Y ~ \forall g \in \mathcal{G}$.

\end{itemize}

The first condition depends only on the $\mathcal G$ group, while the second depends on the specific embedding of $\mathcal{G}$ into $U(3)$.  Different realisations of the same group can generally induce different decompositions of the space $R^\ast\otimes R$, and consequently different $\chi$-covariant subspaces in which $Y$ may reside. Several examples were presented in Appendix~D of Ref.~\cite{Kuncinas:2025uty}. For instance, for the HF-like GOOFy realisation of $S_3$ (for a recent analysis, including the Yukawa sector, see Ref.~\cite{deMedeirosVarzielas:2026svz}), two inequivalent embeddings were identified:
\begin{equation}
S_3 = \left\langle
\begin{pmatrix}
0 & 1 & 0\\
1 & 0 & 0\\
0 & 0 & 1
\end{pmatrix},
\begin{pmatrix}
e^{2 i \pi/3} & 0 & 0\\
0 & e^{-2 i \pi/3} & 0\\
0 & 0 & 1
\end{pmatrix}
\right\rangle,
\end{equation}
results in $\mu_{ij}^2=0$, while
\begin{equation}
S_3' = \left\langle
\begin{pmatrix}
0 & 1 & 0\\
1 & 0 & 0\\
0 & 0 & 1
\end{pmatrix},
\begin{pmatrix}
0 & e^{-2 i \pi/3} & 0\\
e^{2 i \pi/3} & 0 & 0\\
0 & 0 & 1
\end{pmatrix}
\right\rangle,
\end{equation}
yields $\mu_{22}^2=-\mu_{11}^2$.

The above two conditions have immediate consequences for simple groups. For any non-Abelian simple group $\mathcal G$, the commutator subgroup satisfies $[\mathcal G,\mathcal G]=\mathcal G$, and hence the Abelianisation is trivial, $\mathcal G/[\mathcal G,\mathcal G]=\{e\}$. As a result, no non-trivial one-dimensional character exists. The only simple group admitting a non-trivial $\mathbb Z_2$ character is $\mathbb Z_2$ itself.

This conclusion applies to simple groups considered in isolation. In composite groups, however, a $\mathbb Z_2$ factor from another subgroup may supply the required character, allowing for a non-trivial implementation of $V_2$ with a GOOFy transformation.

The preceding discussion assumes $\mathcal X= -\mathcal I$. By contrast, a partial $\mathcal X$ anomaly cannot be extracted as a simple scalar; instead, it fractures the unitary representation $R(g)$. This motivates the next subsection, where we specialise to even cyclic groups and analyse how their sign representation organises the admissible realisations.

\subsection[Orbits of \texorpdfstring{$\mathbb Z_{2m}$}{Z2m}]{Orbits of \boldmath$\mathbb Z_{2m}$}

Consider a cyclic symmetry group
\begin{equation}
\mathcal G=\mathbb{Z}_n=\langle g \mid g^n=e\rangle.
\end{equation}
Since $\mathcal G$ is Abelian, all of its irreducible representations are one-dimensional and are given by the characters
\begin{equation}
\rho_k(g^\ell) = e^{2\pi i k\ell / n}, \quad \ell \in \mathbb Z_n.
\end{equation}
The tensor product of two irreducible representations is given by
\begin{equation}
    \rho_k\otimes\rho_\ell
    \simeq
    \rho_{k+\ell\;(\mathrm{mod}\;n)},
\end{equation}
while complex conjugation acts as
\begin{equation}
    \rho_k^\ast
    \simeq
    \rho_{-k}
    \equiv
    \rho_{n-k}.
\end{equation}

For $\mathbb Z_{2m}$ there exists a sign representation, $\chi = \rho_m$, which satisfies:
\begin{equation}
\chi(g)=-1,\qquad \chi^2=\rho_0.
\end{equation}
Together with the trivial representation $\rho_0$, these are the only self-conjugate irreducible representations, $\rho_k^\ast=\rho_k$ for $k =  \{0,\,m\}$. All remaining irreducible representations occur in complex-conjugate pairs $\{\rho_k,\, \rho_k^\ast = \rho_{2m-k} \}$, for $k= \overline{1,\,m-1}$. 

For $ \mathbb Z_{2m}$ the sign representation acts on the irreducible representations by
\begin{equation}
\rho_k\otimes\chi \simeq \rho_{k+m},\,(\mathrm{mod}\;2m),
\end{equation}
(\textit{i.e.}, every conjugated irreducible representation is shifted by one half of the cyclic group $\rho_k \mapsto \rho_k^\ast \otimes \chi = \rho_{-k + m}$) which partitions the irreducible representations into pairs
\begin{subequations}
\begin{align}\label{Eq:O_sign_pairs}
\mathcal O_0 ={}& \{\rho_0,\rho_m\} = \{1,-1\},\\
\mathcal O_k ={}& \{\rho_k,\rho_{k+m}\},  \qquad\qquad k= \overline{1,\,m-1}.
\end{align}
\end{subequations}

Note that for $\mathbb Z_{2m+1}$ only the trivial representation is self-conjugate, $\rho_0^\ast=\rho_0$, while all remaining irreducible representations form conjugated pairs. Unlike for $\mathbb{Z}_{2m}$, there is no sign representation, and hence no pairing $\rho_k \leftrightarrow \rho_{k+m}$.

The orbit method classifies basis-equivalence classes of symmetry embeddings, but it does not establish a one-to-one correspondence between a chosen generator and a unique set of surviving bilinear couplings. We have already encountered this behaviour in the previous subsection in the two inequivalent embeddings of the $S_3$-symmetric 3HDM, which lead to different $V_2$ sectors.

A particularly instructive example is provided by the $\mathbb{Z}_8$-symmetric 4HDM. This represents the lowest-order realisable cyclic symmetry for which multiple non-trivial sign-paired orbits can be populated independently without inducing algebraic dependence among the doublet representations. Indeed, multiplication by the sign representation $\chi$ partitions the irreducible representations of $\mathbb{Z}_8$ into four disjoint sign-orbits. Different embeddings correspond to different assignments of the scalar doublets among these orbit classes. However, the orbit assignment alone does not determine the surviving bilinear terms. These are obtained only after imposing the GOOFy selection rule that a bilinear may connect two fields whose irreducible representations differ by the sign character.

In general, for $N$ scalar doublets, the potential is constrained by $N-1$ independent relative phases due to the overall $U(1)_Y$ invariance. The maximal realisable cyclic symmetry that can be imposed on this potential without inducing accidental continuous symmetries is $\mathbb{Z}_{2^{N-1}}$. For this, the irreducible representations can be grouped into $2^{N-2}$ sign-orbits. Because the number of sign-orbits $2^{N-2}$ grows exponentially while the physical relative phases $N-1$ grow linearly, locking the potential requires a uniquely determined charge assignment to bridge the populated orbits and prevent the cyclic symmetry from being enhanced to a  continuous $U(1)$ group.

As an example, consider the three embeddings of the $\mathbb Z_8$-symmetric 4HDM:
\begin{subequations}
\begin{align}
\mathcal U_A &= \mathrm{diag}(1,\rho,\rho^2,\rho^3),\\
\mathcal U_B &= \mathrm{diag}(1,1,\rho^2,\rho^3),\\
\mathcal U_C &= \mathrm{diag}(1,-1,\rho^2,\rho^3),
\end{align}
\end{subequations}
where $\rho=e^{i\pi/4}$, so that $\rho^4=-1$ and $\rho^8=1$, and let $\mathcal X=-\mathcal I$.

A bilinear coupling is allowed only if the corresponding pair of fields transforms in irreducible representations related by the sign representation, \textit{i.e.}, whose $\mathbb Z_8$ charges differ by $4$ modulo $8$. Since no such pair is present in either $\mathcal U_A$ or $\mathcal U_B$, one finds $V_2(\mathcal U_A)=V_2(\mathcal U_B)=0$. By contrast, in $\mathcal U_C$ the first two doublets carry charges differing by $4$, so that the bilinear $V_2(\mathcal U_C)=\mu_{12}^2h_{12}+\mathrm{h.c.}$ is allowed. Furthermore, these embeddings generate distinct quartic potentials.

Having illustrated the general structure of the orbit construction in the 4HDM, we now return to the 3HDM and list inequivalent embeddings of the $\mathbb{Z}_{2m}$ group:
\begin{subequations}
\begin{align}
\mathbb{Z}_1 :{}& \quad \mathcal{O}^{(3,0)}; \\
\mathbb{Z}_2 :{}& \quad \mathcal{O}^{(3,0)}, \quad \mathcal{O}^{(2,1)}; \\
\mathbb{Z}_4 :{}& \quad \mathcal{O}^{(3,0)}, \quad \mathcal{O}^{(2,1)} , \quad \mathcal{O}^{(2,0)}\,\mathcal{O}^{(1,0)}, \quad \mathcal{O}^{(1,1)}\,\mathcal{O}^{(1,0)}.
\end{align}
\end{subequations}

Here, $\mathcal{O}^{(n_{\rho_k},n_{\rho_{k+m}})}$ denotes a sign-orbit labelled by the multiplicities of the two irreducible representations related by the sign character. Since the orbit is determined only by these multiplicities, ordering does not matter: $\mathcal{O}^{(m,n)} = \mathcal{O}^{(n,m)} $ and $\mathcal{O}_A \mathcal{O}_B = \mathcal{O}_B \mathcal{O}_A$. The orbit notation provides a convenient book-keeping technique for the eigenspace decomposition induced by the sign representation of $\mathcal X$, recording how the scalar doublets are distributed among the corresponding $\pm 1$ eigenspaces.

The inclusion of the trivial group $\mathbb{Z}_1$ serves to identify the origin of the class $\mathcal{O}^{(3,0)}$. Moreover, every $\mathbb{Z}_{2(m+1)}$ embedding contains all orbit classes already present in $\mathbb{Z}_{2m}$, a consequence of the additional global $U(1)$ rephasing freedom of the scalar potential. The remaining orbit classes therefore correspond to new embeddings. In this way, the orbit classification provides a compact description of all realisable cyclic embeddings in the 3HDM and makes explicit which symmetry realisations are distinct up to basis transformations. It also identifies which invariant subspaces already appear in lower-order embeddings and which arise only at higher cyclic order. We use this classification as the starting point for the analysis of the GOOFy 3HDMs, where the orbit structure is extended by the additional sign-flipping sector and the corresponding HF-like and GCP-like realisations.

\section{Different realisable GOOFy cases}\label{Sec:Realisable_GOOFy}

Since the highest realisable cyclic group contains all embeddings of different orbit classes, it is sufficient to adopt it as the reference embedding. The $\mathbb{Z}_4$-symmetric 3HDM yields four inequivalent configurations:
\begin{align*}
&{} \mathcal{O}^{(3,0)}, \quad \mathcal{O}^{(2,1)}, \quad \mathcal{O}^{(2,0)}\,\mathcal{O}^{(1,0)},\quad \mathcal{O}^{(1,1)}\,\mathcal{O}^{(1,0)}.
\end{align*}

A non-trivial quadratic sector can only arise if at least one orbit is fully populated. For instance, choosing $\mathcal{X} = - \mathcal{I}$ we get that the $\mathcal{O}^{(3,0)}$ class does not contain any pairs, implying $V_2=0$. In contrast, the $\mathcal{O}^{(2,1)}$ class contains a paired orbit, and consequently $V_2 \neq 0$. 

A different situation occurs for the  $\mathcal X = \mathrm{diag}(-1,\, -1,\, 1)$ assignment, for which one of the scalar doublets transforms trivially under the sign representation. Consequently, the orbit $\mathcal{O}^{(3,0)}$ admits $V_2 = \mu_{33}^2 h_{33}$ since the $h_3$ sector is invariant under the action of $\mathcal X$.

\subsection{The overall sign-flipping GOOFy transformations}\label{Sec:Overall_X_discussion}

In Ref.~\cite{Kuncinas:2025uty} several HF-like GOOFy transformations were noted that do not automatically lead to a scalar potential with a vanishing quadratic sector:\newpage 
\begin{align}
& V_2 = \mu_{11}^2 (h_{11} - h_{22}): && \{\mathbb{Z}_4,\, D_4,\, S_3,\, O(2),\, O(2) \times U(1),\, U(1) \circ V_4\},\\
& V_2 = \mu_{12}^2 h_{12} + \mathrm{h.c.}: &&\{\mathbb{Z}_2 \times \mathbb{Z}_2,\, U(1) \times \mathbb{Z}_2\},\\
& V_2 = \mu_{12}^2 h_{12} + \mu_{13}^2 h_{13} + \mathrm{h.c.}: && \{\mathbb{Z}_2\}.
\end{align}
Although $V_2$ take different forms, all these cases have $\mathrm{tr}(Y)=0$ with the eigenvalues given by $\{-\lambda_a, 0, \lambda_a\}$. Therefore, there exists a basis where $V_2$ takes a common form.

At this stage, it is important to identify the residual basis freedom; see Ref.~\cite{Kuncinas:2025uty} for a general discussion in the 3HDM. Let $\mathcal{U} = \mathrm{diag}(a_1,\,a_2,\,a_3)$. Its centraliser in $SU(3)$ (and, up to isomorphism, in $PSU(3)$) consists of all unitary transformations preserving the eigenspace decomposition of $\mathcal{U}$. The centraliser depends only on the multiplicities of the eigenvalues and is therefore invariant under $\mathcal{U} \to -\mathcal{U}$. Three possibilities arise:
\begin{equation}
C(\mathcal{U}) \cong
\begin{cases}
S(U(1)^3)\cong U(1)^2, & a_i \neq a_j \text{ for } i \neq j,\\[2mm]
S(U(2)\times U(1))\cong U(2), & a_i = a_j \neq a_k,\\[2mm]
SU(3), & a_1 = a_2 = a_3.
\end{cases}
\end{equation}
The centraliser determines the residual basis freedom and therefore the number of redundant parameters in the scalar potential.

As an illustration, the centraliser of the $\mathbb{Z}_2$ generator is $U(1) \times U(2)$. After factoring out the overall hypercharge phase, four redundant degrees of freedom remain. These can be used to eliminate one complex off-diagonal coupling, render the remaining one real, and subsequently rotate all equivalent quadratic potentials into a common form. Although a general $U(3)$ basis transformation can always diagonalise $Y$, such a basis does not, in general, make the underlying symmetry manifest.

This basis reduction is consistent with the orbit classification of the previous subsection: the orbit structure determines which bilinears can appear, while the centraliser determines which of those forms are related by residual basis transformations.

The anti-commutation property $Y \mathcal{U} = -\mathcal{U} Y$ dictates (note that we are discussing $\mathcal X = -\mathcal I$) the eigenvalue spectrum of the quadratic matrix. If $m^2$ is an eigenvalue of $Y$ with eigenvector $v$, the invariance condition yields:
\begin{equation}
Y (\mathcal{U} v) = -m^2 (\mathcal{U} v).
\end{equation}
Therefore, the non-zero eigenvalues of $Y$ must appear in symmetric pairs, $\pm m^2$. The consequences for an NHDM depend intrinsically on the parity of $N$:
\begin{itemize}
\item $N$ is even: any non-zero eigenvalue must be accompanied by its negative;
\item $N$ is odd: the eigenvalues cannot all be paired, this forces at least one zero mode.
\end{itemize}
These general constraints will be illustrated in the 4HDM and 5HDM, where all admissible eigenvalue patterns are listed; see eqs.~\eqref{Eq:4HDM_V2_HF_eigenvalues}--\eqref{Eq:5HDM_V2_GCP_eigenvalues}.

Moving on to the the GCP-like GOOFy transformations, the interesting cases of Ref.~\cite{Kuncinas:2025uty} are: $ \{ \mathbb{Z}_2^\ast\, ,\mathbb{Z}_4^\ast,\, \mathbb{Z}_2 \times \mathbb{Z}_2^\ast, \, \mathbb{Z}_3 \rtimes \mathbb{Z}_2^\ast, \, D_4, \, U(1) \rtimes \mathbb{Z}_2^\ast, \, U(1)_2 \rtimes \mathbb{Z}_2^\ast, \, SO(2) \rtimes \mathbb{Z}_2^\ast$, \linebreak  $([U(1) \times D_4]/ \mathbb{Z}_2) \rtimes \mathbb{Z}_2^\ast \} $. Consistent with the HF-like transformations, in all these cases the quadratic part of the potential can be rotated into a basis with $V_2 = \mu_{11}^2 (h_{11} - h_{22})$.

Rather than treating each composite symmetry separately, it is more instructive to identify the elementary sign-flipping GOOFy generators from which all of the above cases are constructed. Up to basis transformations, the HF-like overall sign-flipping GOOFy transformations are generated by (we denote GOOFy groups by a tilde, while the subscript ``$-$'' indicates the overall sign flip):
\begin{itemize}
\item $\tilde{\mathbb{Z}}_{1,-}  =  \mathcal{W}\,(\mathcal{I}, -\mathcal{I}, 0)$

This case renders $V_2=0$, while $V_4$ is the most general one. Note that $\mathbb{Z}_1 = \mathcal{I}$. In the extended $H$ space the $6\times6$ matrix spans $\mathbb{Z}_2$;

\item $\tilde{\mathbb{Z}}_{2,-} =  \mathcal{W}\,(\mathbb{Z}_2, -\mathcal{I}, 0)$

Choosing the generator $\mathbb{Z}_2 = \mathrm{diag}(1,\,1,\,-1)$ yields the quadratic potential:
\begin{equation}
V_2 = \mu_{13}^2 h_{13} + \mu_{23}^2 h_{23} + \mathrm{h.c.},
\end{equation}
and the quartic potential:
\begin{equation}
\begin{split}
V_4 ={}& \sum_i \lambda_{iiii} h_{ii}^2 + \sum_{i<j} \lambda_{iijj} h_{ii} h_{jj} + \sum_{i<j} \lambda_{ijji} h_{ij} h_{ji}\\
&  + \Big\lbrace \sum_{i<j} \lambda_{ijij} h_{ij}^2 + \lambda_{1112} h_{11}h_{12} + \lambda_{1222} h_{12}h_{22}\\
& \qquad + \lambda_{1233} h_{12}h_{33} + \lambda_{1323} h_{13}h_{23} + \lambda_{1332} h_{13}h_{32} + \mathrm{h.c.} \Big\rbrace;
\end{split}
\end{equation} 

\item $\tilde{\mathbb{Z}}_{4,-}  =  \mathcal{W}\,(\mathbb{Z}_4, -\mathcal{I}, 0)$

Choosing the generator $\mathbb{Z}_4 = \mathrm{diag}(1,\,-1,\,i)$ constrains the quadratic sector to:
\begin{equation}
V_2 = \mu_{12}^2 h_{12} + \mathrm{h.c.},
\end{equation}
while the quartic potential is given by:
\begin{equation}
\begin{aligned}
V_4 ={}& \sum_i \lambda_{iiii} h_{ii}^2 + \sum_{i<j} \lambda_{iijj} h_{ii} h_{jj} + \sum_{i<j} \lambda_{ijji} h_{ij} h_{ji}\\
&  + \Big\lbrace  \lambda_{1212} h_{12}^2 + \lambda_{1323} h_{13}h_{23} + \mathrm{h.c.}  \Big\rbrace;
\end{aligned}
\end{equation} 
\end{itemize}
while the unique GCP-like overall sign-flipping GOOFy transformations are (the even order of GCP transformations stems from their anti-linear (or anti-unitary) nature, as they involve complex conjugation):
\begin{itemize}
\item $\tilde{\mathbb{Z}}_{2,-}^\ast =  \mathcal{W}\,(\mathbb{Z}_2, -\mathcal{I}, 1)$

In the $H$ space the $6\times6$ matrix spans $\mathbb{Z}_4/\left\langle -\mathcal{I}_6 \right\rangle \cong \mathbb{Z}_2$.

A convenient basis choice would be to require invariance under $\mathcal{W}\,(\mathcal{I}, -\mathcal{I}, 1)$. This trivial realisation yields the quadratic potential
\begin{equation}
V_2 = i \sum_{i<j} \mu_{ij}^2 h_{ij}  + \mathrm{h.c.},
\end{equation}
accompanied by the most general quartic potential with purely real couplings.

An equivalent choice of $\mathcal{W}\,(\mathbb{Z}_2, -\mathcal{I}, 1)$ would render some of the quadratic couplings real, while the corresponding quartic couplings would become purely imaginary;

\item $\tilde{\mathbb{Z}}_{4,-}^\ast =  \mathcal{W}\,(\mathbb{Z}_4, -\mathcal{I}, 1)$

 A choice of the generator
\begin{equation}
\mathbb{Z}_4^\ast = \begin{pmatrix}
0 & -1 & 0 \\
1 & 0 & 0 \\
0 & 0 & i
\end{pmatrix},
\end{equation}
constrains the quadratic sector to
\begin{equation}
V_2 = \mu_{11}^2 (h_{11} - h_{22}) +  \left\lbrace \mu_{12}^2 h_{12} + \mathrm{h.c.} \right\rbrace,
\end{equation}
and results in the following quartic potential:
\begin{equation}
\begin{aligned}
V_4={}& \lambda_{1111} (h_{11}^2 + h_{22}^2) + \lambda_{3333} h_{33}^2 + \lambda_{1221} h_{12}h_{21} + \lambda_{1331} (h_{13}h_{31} + h_{23}h_{32}) \\
& + \lambda_{1122} h_{11}h_{22} + \lambda_{1133} (h_{11} + h_{22}) h_{33} + \lambda_{1323} (h_{13}h_{23} + h_{31}h_{32}) \\
& + \left\lbrace \lambda_{1212} h_{12}^2 + \lambda_{1313}(h_{13}^2-h_{32}^2)  - \lambda_{1112} h_{12}(h_{11} - h_{22}) + \mathrm{h.c.}\right\rbrace.
\end{aligned}
\end{equation}
\end{itemize}

In terms of the orbit decomposition, the GOOFy constructions are classified by
\begin{equation*}
\tilde{\mathbb{Z}}_1\,\left[\mathcal{O}^{(3,0)}\right],\qquad
\tilde{\mathbb{Z}}_2\,\left[\mathcal{O}^{(2,1)}\right],\qquad
\tilde{\mathbb{Z}}_4\,\left[\mathcal{O}^{(1,1)}\,\mathcal{O}^{(1,0)}\right].
\end{equation*}
The only orbit configuration not listed is $\tilde{\mathbb{Z}}_4\,[\mathcal{O}^{(2,0)}\,\mathcal{O}^{(1,0)}]$: neither orbit contains a fully populated pair of irreducible representations. Consequently, no invariant bilinear can be constructed and the quadratic sector necessarily vanishes, $V_2=0$. As a result, $\tilde{\mathbb{Z}}_4\,[\mathcal{O}^{(2,0)}\,\mathcal{O}^{(1,0)}]$ corresponds to $\mathbb{Z}_4 \times \tilde{\mathbb{Z}}_1\,\left[\mathcal{O}^{(3,0)}\right]$.

All other symmetry groups are reducible to one of the above five generators and with the conventional HF or GCP symmetry applied on top. For instance, a $\mathbb{Z}_3$ GOOFy transformation has no homomorphism $\varphi: \mathbb{Z}_n \to \mathbb{Z}_2$, see Subsection~\ref{Sec:Reps_mismatch}, and hence yields $V_2 = 0$, thereby mapping to $\tilde{\mathbb{Z}}_{1,-}$. Thus, the list of the five generators provides a complete basis for the distinct classes imposed by the overall sign-flipping GOOFy transformations. 

To understand the viability of the above models, we must evaluate their stationary-point equations. We denote the most general vacuum expectation value (vev) alignment as $(v_1,\, v_2,\, v_3) \in \mathbb{C}$. The non-Abelian groups possess multi-dimensional irreducible representations, meaning the Lagrangian cannot distinguish between fields within the same multiplet. This results in vacuum configurations with relations between the vevs, \textit{e.g.}, fixing the ratio $v_i / v_j$. On the other hand, $\mathbb{Z}_n$ symmetries possess strictly one-dimensional irreducible representations. Because the doublets transform independently, the Lagrangian distinguishes between them. Consequently, the stationary-point equations yield either entirely generic, continuous vevs or vanishing entries. This strictly Abelian behaviour significantly simplifies the vacua one has to consider.

Furthermore, if two vacua are related by a trivial permutation of indices (\textit{e.g.}, swapping $h_i \leftrightarrow h_j$ within a subspace where the applied symmetry treats them identically), the resulting physics is equivalent up to a basis transformation. Consequently, we do not need to consider these redundant copies.

In Table~\ref{Table:Vacua_overall_X} we present only non-equivalent vacua. As summarised in Table~\ref{Table:Vacua_overall_X}, extracting the scalar mass spectrum reveals that in many cases there are massless states or negative mass-squared eigenvalues (saddle points) present; these instances are highlighted in grey. We recall that the overall sign-flipping GOOFy cases differ only in their quadratic couplings. Consequently, introducing soft terms reduces these to the usual symmetries.

{\renewcommand{\arraystretch}{1.35}
\begin{table}[H]
\caption{Vacua and scalar mass spectrum classifications for overall sign-flipping GOOFy transformations. Here, $m_H = 0$ indicates the presence of a massless scalar, ``$m_H$ saddle" denotes a saddle point ($m_H^2 < 0$), ``$m_H$ degen." indicates mass-degenerate scalar states, and ``$\mathbb{R}_\text{vev}$" specifies that massless scalars emerge exclusively for real vevs. Non-viable implementations are highlighted in grey. Explicit (trivial) parameter constraints resulting from the stationary-point equations are displayed where applicable; however, in none of these cases are the underlying conditions found to be RG-stable. Empty cells denote configurations that yield massive spectra without unique parameter relations.}
\label{Table:Vacua_overall_X}
\begin{center}
\begin{tabular}{|c||c|c|c||c|c|} \hline\hline

& \multicolumn{3}{c||}{HF-like} & \multicolumn{2}{c|}{GCP-like} \\ \cline{2-6} 
& $\tilde{\mathbb{Z}}_{1,-}$ & $\tilde{\mathbb{Z}}_{2,-}$ & $\tilde{\mathbb{Z}}_{4,-}$ & $\tilde{\mathbb{Z}}_{2,-}^\ast$ & $\tilde{\mathbb{Z}}_{4,-}^\ast$ \\ \hline \hline

$(v_1, v_2, v_3)$ & \cellcolor{gray!18} $m_H = 0$ &  & $\lambda_{1323}^\mathrm{I}=0$  & 
\cellcolor{gray!18} \begin{tabular}[l]{@{}c@{}@{}} $\mathbb{R}_\text{vev}:$ \\ $m_H = 0$  \end{tabular} & $\lambda_{1313}^\mathrm{I}=0$ \\ \hline \hline

$(v_1,v_2,0)$ \rule[-1em]{0pt}{2.2em} &  \multicolumn{1}{c|}{\cellcolor{gray!18}} &  \cellcolor{gray!18} $m_H = 0$ &  &  \multicolumn{1}{c|}{\cellcolor{gray!18}} & 
  \\[-0.1pt] \hhline{-|>{\arrayrulecolor{gray!18}}->{\arrayrulecolor{black}}|-|-|>{\arrayrulecolor{gray!18}}->{\arrayrulecolor{black}}|-|}

$(v_1, 0, v_3)$\rule[-1em]{0pt}{2.2em} &  \multicolumn{1}{c|}{\multirow{-2}{*}{\cellcolor{gray!18}$m_H = 0$}} &  &  \cellcolor{gray!18} $m_H$ saddle &  \multicolumn{1}{c|}{\multirow{-2}{*}{\cellcolor{gray!18}\makecell[c]{$\mathbb{R}_\text{vev}:$ \\ $\mu_{12}^2=0$\\ $m_H=0$}}} &  $\lambda_{1313}^\mathrm{I}=0$ \\ \hline \hline

$(v, 0, 0)$ & 
  \multicolumn{1}{c|}{\cellcolor{gray!18}} & 
  \cellcolor{gray!18}\begin{tabular}[l]{@{}c@{}@{}} $\mu_{13}^2 = 0$ \\ $\lambda_{1111} = \lambda_{1112}=0$ \\ $m_H = 0$ \end{tabular} & 
  \cellcolor{gray!18}\begin{tabular}[l]{@{}c@{}@{}} $V_2=0$ \\ $\lambda_{1111}=0$ \\$m_H = 0$ \\ $m_H$ degen. \end{tabular} & 
  \multicolumn{1}{c|}{\cellcolor{gray!18}} & 
\\[-0.1pt] \hhline{-|>{\arrayrulecolor{gray!18}}->{\arrayrulecolor{black}}|-|-|>{\arrayrulecolor{gray!18}}->{\arrayrulecolor{black}}|-|}

$(0, 0, v)$ & 
  \multicolumn{1}{c|}{\multirow{-4}{*}{\cellcolor{gray!18}\begin{tabular}[l]{@{}c@{}@{}} $\lambda_{1111} = \lambda_{1112}$ \\ $= \lambda_{1113} = 0$ \\ $m_H = 0$ \end{tabular}}} & 
  \cellcolor{gray!18}\begin{tabular}[l]{@{}c@{}@{}} $V_2 = 0$ \\ $\lambda_{3333} = 0$ \\ $m_H = 0$ \end{tabular} & 
  \cellcolor{gray!18}\begin{tabular}[l]{@{}c@{}@{}} $\lambda_{3333} = 0$ \\ $m_H = 0$ \end{tabular} & 
  \multicolumn{1}{c|}{\multirow{-4}{*}{\cellcolor{gray!18}\begin{tabular}[l]{@{}c@{}@{}} $\mu_{12}^2 = \mu_{13}^2 = 0 $ \\ $\lambda_{1111} = \lambda_{1112}$ \\ $= \lambda_{1113} = 0$ \\ $m_H = 0$ \end{tabular}}} & 
  \cellcolor{gray!18}\begin{tabular}[l]{@{}c@{}@{}} $\lambda_{3333} = 0$ \\ $m_H = 0$ \end{tabular} \\ \hline\hline

\end{tabular}
\end{center}
\end{table}}

\subsection{The partial sign-flipping GOOFy transformations}\label{Sec:Partial_X_discussion}

Building on the overall sign-flipping scenario discussed in the previous subsection, we now consider the more general framework where $\mathcal{X}$ acts non-trivially only on a specific subspace of the multiplet $H$.

With the developed tools we can resolve an apparent ambiguity noted in Ref.~\cite{Kuncinas:2025uty}: how the same $V_4$ structure can correspond to two distinct $V_2 \neq 0$ structures. When $\mathcal{X}$ acts non-uniformly, the $N$-dimensional field multiplet splits into two distinct blocks: one anomalous and one invariant. By choosing between $\mathbf{1}$ and $\mathbf{1^\prime}$ of $\mathcal X$ (effectively swapping which block is anomalous) the $V_2$ part yields two distinct, yet complementary, quadratic potentials for the identical $V_4$ structure.

To systematically identify the seeds for the partial sign-flipping GOOFy transformations, we recall from Subsection~\ref{Sec:Overall_X_discussion} that the basic building blocks were given by $\mathcal{U} \in \{\mathbb{Z}_1,\,\mathbb{Z}_2,\,\mathbb{Z}_4\}$. It should be of no surprise that in the partial sign-flipping GOOFy framework we reproduce five unique overall sign-flipping transformations; after all, this splitting is dictated by the different sign-orbit classes. However, in the current transformations the bilinear $SU(2)$ singlets $h_{ij}$ no longer transform as even quantities, leading to some changes in the $V_4$ part of the potential.

To simplify notation, we shall fix:
\begin{equation}
\mathcal{X}^\prime = \mathrm{diag}(-1,\,-1,\,1).
\end{equation}

For the HF-like partial sign-flipping GOOFy transformations we identify the following basic blocks (note that the scalar potentials are presented in the invariant form, without the further simplification of the redundant couplings):
\begin{itemize}
\item $\tilde{\mathbb{Z}}_{1,\pm}  =  \mathcal{W}\,(\mathcal{I},\, \pm \mathcal{X}^\prime,\, 0)$

The quartic scalar potential is given by:
\begin{equation}
\begin{aligned}
V_4 = {}& \lambda_{1111} h_{11}^2 + \lambda_{2222} h_{22}^2 + \lambda_{3333} h_{33}^2 + \lambda_{1221} h_{12} h_{21} + \lambda_{1122}h_{11}h_{22} \\
&+ \bigg\{ \lambda_{1112} h_{11}h_{12} + \lambda_{1222} h_{12}h_{22} + \lambda_{1323} h_{13}h_{23} \\
&\qquad+ \lambda_{1212} h_{12}^2 + \lambda_{1313} h_{13}^2 + \lambda_{2323} h_{23}^2 + \mathrm{h.c.} \bigg\},
\end{aligned}
\end{equation}
while the quadratic part is given by either
\begin{equation}
V_{2,\mathcal{X}^\prime} = \mu_{33}^2 h_{33},
\end{equation}
or
\begin{equation}
V_{2,-\mathcal{X}^\prime} = \mu_{11}^2 h_{11} + \mu_{22}^2 h_{22} + \left\lbrace \mu_{12}^2 h_{12} + \mathrm{h.c.} \right\rbrace.
\end{equation}

These potentials were presented as $\mathcal{G}_2$ and $\mathcal{G}_1$ in Ref.~\cite{Kuncinas:2025uty} respectively;

\item $\tilde{\mathbb{Z}}_{2,\pm} =  \mathcal{W}\,(\mathrm{diag}(1,-1,-1),\, \pm\mathcal{X}^\prime,\, 0)$

\begin{equation}
\begin{aligned}
V_4 = {}& \lambda_{1111} h_{11}^2 + \lambda_{2222} h_{22}^2 + \lambda_{3333} h_{33}^2 + \lambda_{1221} h_{12} h_{21} + \lambda_{1122}h_{11}h_{22}\\
& + \left\lbrace \lambda_{1332} h_{13}h_{32} + \lambda_{1233} h_{12}h_{33} + \lambda_{1212} h_{12}^2 + \lambda_{1313} h_{13}^2 + \lambda_{2323} h_{23}^2 + \mathrm{h.c.} \right\rbrace,
\end{aligned}
\end{equation}
while the quadratic part is given by either
\begin{equation}
V_{2,\mathcal{X}^\prime} = \left\lbrace \mu_{12}^2 h_{12} + \mathrm{h.c.} \right\rbrace + \mu_{33}^2 h_{33},
\end{equation}
or
\begin{equation}
V_{2,-\mathcal{X}^\prime} = \mu_{11}^2 h_{11} + \mu_{22}^2 h_{22}.
\end{equation}

This case corresponds to $\mathcal{G}_3$ of Ref.~\cite{Kuncinas:2025uty};

\item $ \tilde{\mathbb{Z}}_{4,\pm} =  \mathcal{W}\,(\mathrm{diag}(1,-1,i),\, \pm\mathcal{X}^\prime,\, 0)$
\begin{equation}
\begin{aligned}
V_4 = {}&  \lambda_{1111} h_{11}^2 + \lambda_{2222} h_{22}^2 + \lambda_{3333} h_{33}^2 + \lambda_{1221} h_{12} h_{21} + \lambda_{1122}h_{11}h_{22}\\
&  + \left\lbrace \lambda_{1323} h_{13}h_{23} + \lambda_{1332} h_{13}h_{32} + \lambda_{1233} h_{12}h_{33} + \lambda_{1212} h_{12}^2 + \mathrm{h.c.} \right\rbrace.
\end{aligned}
\end{equation}
while the quadratic part is given by either
\begin{equation}
V_{2,\mathcal{X}^\prime} = \left\lbrace \mu_{12}^2 h_{12} + \mathrm{h.c.} \right\rbrace + \mu_{33}^2 h_{33},
\end{equation}
or
\begin{equation}
V_{2,-\mathcal{X}^\prime} = \mu_{11}^2 h_{11} + \mu_{22}^2 h_{22}.
\end{equation}

This case corresponds to $\mathcal{G}_5$ of Ref.~\cite{Kuncinas:2025uty}.

\end{itemize}

We note that the definition of $\mathcal{G}_4$ in Ref.~\cite{Kuncinas:2025uty} is incorrect, as it fails to yield $\mathcal{X}$ with eigenvalues of $\pm1$. This demonstrates the advantage of utilising the $\mathcal{W}$ transformations of eq.~\eqref{Eq:Def_W_GOOFy_U_X} with a diagonal $\mathcal X$.

There is an additional subtlety when $\mathcal{X}$ is not proportional to the identity. The symmetry exhibits an interchange of the two blocks, $
\mathcal U \leftrightarrow \mathcal U^\ast\mathcal X$, which leads to different generators (in the $3 \times 3$ space) describing the same scalar potential. For example, imposing invariance under $\mathcal{W}\,(\mathcal I,\mathcal X',0)$ is equivalent, up to the overall transformation $-\mathcal I_6$, to imposing invariance under $\mathcal{W}\,(-\mathcal X',\mathcal X',0)$. The ambiguity originates from the fact that the action of $\mathcal X$ is no longer uniform on all scalar doublets.

To distinguish inequivalent realisations, it is convenient to associate to each generator the pair of traces $\left(\mathrm{tr}\,\mathcal (U),\,\mathrm{tr}\,\mathcal (X)\right)$. Generators with different trace pairs necessarily belong to different equivalence classes, whereas generators sharing the same trace pair require a further comparison under residual basis transformations.

This observation also indicates that the sign-orbit classification introduced previously requires an additional grading whenever $\mathcal X$ is non-uniform. Besides the orbit multiplicities, one must specify how the individual irreducible representations transform under the sign representation. For instance, the generators
\begin{equation*}
\mathcal W\,(\mathcal I,\mathcal X',0),
\qquad
\mathcal W\,(\mathcal I,-\mathcal X',0),
\end{equation*}
may be represented schematically by
\begin{equation*}
\mathcal O^{(2,0)}\mathcal S^{(1)},
\qquad
\mathcal O^{(1,0)}\mathcal S^{(2)},
\end{equation*}
to indicate, respectively, the orbit structure together with the number of the $\mathcal X$-singlet fields, $S$. This additional grading is needed because the orbit decomposition alone does not fully resolve the inequivalent sign assignments when \(\mathcal X\) is non-uniform.

Moving on, for the GCP-like partial sign-flipping GOOFy transformations we identify the following blocks:
\begin{itemize}
\item $ \tilde{\mathbb{Z}}_{2,\pm}^\ast  =  \mathcal{W}\,(\mathcal{I},\, \pm \mathcal{X}^\prime,\, 1)$

The quartic scalar potential is given by:
\begin{equation}
\begin{aligned}
V_4 = {}& \lambda_{1111} h_{11}^2 + \lambda_{2222} h_{22}^2 + \lambda_{3333} h_{33}^2 + \lambda_{1221} h_{12} h_{21} + \lambda_{1122}h_{11}h_{22} \\
&+ \bigg\{ \lambda_{1112} h_{11}h_{12} + \lambda_{1222} h_{12}h_{22} + \lambda_{1323} h_{13}h_{23} \\
&\qquad+ \lambda_{1212} h_{12}^2 + \lambda_{1313} h_{13}^2 + \lambda_{2323} h_{23}^2\\
&\qquad- i \lambda_{1332} h_{13}h_{32} + i\lambda_{1233} h_{12}h_{33} + \mathrm{h.c.} \bigg\},
\end{aligned}
\end{equation}
where all couplings are real. The quadratic part is given by either
\begin{equation}
V_{2,\mathcal{X}^\prime} = i \mu_{12}^2 (h_{12} -h_{21}) + \mu_{33}^2 h_{33},
\end{equation}
or
\begin{equation}
V_{2,-\mathcal{X}^\prime} = \mu_{11}^2 h_{11} + \mu_{22}^2 h_{22} + \mu_{12}^2 (h_{12} + h_{21}).
\end{equation}

Note that the choice of either $\mathcal{U}= \{\mathcal{I},\, \mathrm{diag}(1,\,-1,\,i),\, \mathrm{diag}(1,\,-1,\,-1)\}$ (these are the generators of the unique HF-like partial sign-flipping GOOFy transformations) results in the same scalar potential written in a different basis. This can be easily verified by considering the eigenvalues of the $\mathcal{W}$ transformation;

\item $ \tilde{\mathbb{Z}}_{4,\pm}^\ast =  \mathcal{W}\,(\begin{pmatrix}
0 & 1 & 0\\
-1 & 0 & 0\\
0 & 0 & i
\end{pmatrix},\, \pm\mathcal{X}^\prime,\, 1)$

The quartic scalar potential is given by:
\begin{equation}
\begin{aligned}
V_4 = {}& \lambda_{1111} (h_{11}^2 + h_{22}^2) + \lambda_{3333} h_{33}^2 + \lambda_{1221} h_{12} h_{21} + \lambda_{1122}h_{11}h_{22}\\
&  + \lambda_{1133} (h_{22} - h_{11})h_{33} +  \lambda_{1331} (h_{13}h_{31} - h_{23}h_{32}) + \lambda_{1323} (h_{13}h_{23} + h_{31}h_{32})\\
& + \bigg\{ \lambda_{1212} h_{12}^2 + \lambda_{2331} h_{23} h_{31} + \lambda_{1112} (h_{11} - h_{22})h_{21}\\
& \qquad+  \lambda_{1233} h_{12}h_{33} + \lambda_{1313} (h_{13}^2 - h_{32}^2) + \mathrm{h.c.} \bigg\},
\end{aligned}
\end{equation}
while the quadratic part is given by either
\begin{equation}
V_{2,\mathcal{X}^\prime} =  \mu_{11}^2 (h_{11} - h_{22})  + \left\lbrace \mu_{12}^2 h_{12} + \mathrm{h.c.} \right\rbrace + \mu_{33}^2 h_{33},
\end{equation}
or
\begin{equation}
V_{2,-\mathcal{X}^\prime} = \mu_{11}^2 (h_{11} + h_{22}) .
\end{equation}

This case corresponds to $\mathcal{G}_7^\ast$ of Ref.~\cite{Kuncinas:2025uty}.

\end{itemize}

The above GCP-like cases yield $\left( \mathcal{W} \, \mathcal{W}^\ast \right)^2 = \mathcal{I}$.

As noted in Ref.~\cite{Kuncinas:2025uty}, the bilinear spectra of the GOOFy cases exhibit a characteristic feature absent in the ordinary HF or GCP invariant potentials: some eigenvalues occur in pairs with identical absolute value but opposite signs. Up to permutations, $\pm \mathcal X^\prime$ induces the scalar space decomposition into the $\pm1$ eigenspaces, $V=V_+\oplus V_-$. This splits the three scalar doublets into one of the three admissible orbit configurations,
\begin{equation}
\mathcal O^{(2,0)}\mathcal S^{(1)},\qquad
\mathcal O^{(1,1)}\mathcal S^{(1)},\qquad
\mathcal O^{(1,0)}\mathcal S^{(2)}.
\end{equation}
These are the only ways of distributing three doublets among the $\mathcal O^{(m)} \mathcal S^{(n)}$ sectors, subject to the condition $m+n=3$. In our notation, the splitting always occurs between the $\{h_1,h_2\}{-}\{h_3\}$ sectors. This is already visible at the level of the phase-invariant part of the potential, which is governed by
\begin{equation}
V_{U(1)\times U(1)} = \sum_{i=1}^3 \mu_{ii}^2 h_{ii} + \sum_{i=1}^3 \lambda_{iiii} h_{ii}^2 + \sum_{i<j}^3 \lambda_{iijj} h_{ii}h_{jj} + \sum_{i<j}^3 \lambda_{ijji} h_{ij}h_{ji}.
\end{equation}

The mixed bilinears $\{h_{13},\,h_{23}\}$ in the adjoint bilinear space correspond to the $r_{4 \ldots 7}$ entries. Specifically, the paired eigenvalues with equal absolute values appear within this block. In this sense, the property of the absolute magnitudes of the eigenvalues is a direct consequence of the orbit structure induced by $\mathcal X$.

Having identified all possible partial sign-flipping GOOFy realisations in the 3HDM, we now proceed to the analysis of their scalar spectra. The spectrum for each model is presented in Tables~\ref{Table:Vacua_diff_sgn_HF} and \ref{Table:Vacua_diff_sgn_GCP} to facilitate comparison between vacua and models. 

While evaluating the mass spectrum was straightforward for the overall sign-flipping cases, the partial sign-flipping scenarios require further inspection: in Subsection~\ref{Sec:Evaluating_RG} we demonstrate that these configurations are RG-unstable. Therefore, the listed implementations should be interpreted with caution.

{\renewcommand{\arraystretch}{1.4}
\begin{table}[H]
\caption{Classification of vacuum configurations for partial sign-flipping HF-like GOOFy transformations. Notation follows Table~\ref{Table:Vacua_overall_X}.}
\label{Table:Vacua_diff_sgn_HF}
\footnotesize
\begin{center}
\begin{tabular}{|c||c|c||c|c||c|c|} \hline\hline

 & $\tilde{\mathbb{Z}}_1(\mathcal X^\prime)$ & $\tilde{\mathbb{Z}}_1(-\mathcal X^\prime)$ & $\tilde{\mathbb{Z}}_2(\mathcal X^\prime)$ & $\tilde{\mathbb{Z}}_2(-\mathcal X^\prime)$ & $\tilde{\mathbb{Z}}_4(\mathcal X^\prime)$ & $\tilde{\mathbb{Z}}_4(-\mathcal X^\prime)$ \\ \hline \hline

$(v_1, v_2, v_3)$ & & & & & $\lambda_{1323}^\mathrm{I}=0$ & $\lambda_{1323}^\mathrm{I}=0$  \\ \hline\hline

$(v_1, v_2, 0)$ & \cellcolor{gray!18} $m_{H} = 0$ & \cellcolor{gray!18} \begin{tabular}[l]{@{}c@{}@{}} $m_{H^\pm} = 0$ \\ $m_H \text{ saddle}$ \end{tabular} & & $\lambda_{1212}^\mathrm{I} = 0$ & & $\lambda_{1212}^\mathrm{I}=0$\\ \hline

$(v_1, 0, v_3)$ & $\lambda_{1313}^\mathrm{I}=0$ & $\lambda_{1313}^\mathrm{I}=0$ & $\lambda_{1313}^\mathrm{I}=0$  & \begin{tabular}[l]{@{}c@{}@{}} $\lambda_{1313}^\mathrm{I} = 0$ \\ $\lambda_{1332} = -\lambda_{1233}$ \end{tabular} & \cellcolor{gray!18}\begin{tabular}[l]{@{}c@{}@{}} $\lambda_{1111}=0$ \\ $m_H, m_{H^\pm}$ \\ saddle \end{tabular} & \cellcolor{gray!18}\begin{tabular}[l]{@{}c@{}@{}} $\lambda_{3333}=0$ \\ $m_H, m_{H^\pm}$ \\ saddle \end{tabular} \\ \hline \hline

$(v, 0, 0)$ & \cellcolor{gray!18}\begin{tabular}[l]{@{}c@{}@{}}$\lambda_{1111} = 0$ \\ $\lambda_{1112} = 0$ \\ $m_H = 0$ \end{tabular} & \cellcolor{gray!18}\begin{tabular}[l]{@{}c@{}@{}}$m_{H^\pm} = 0$ \\ $m_H \text{ saddle}$ \end{tabular} & \cellcolor{gray!18}\begin{tabular}[l]{@{}c@{}@{}}$\mu_{12}^2 = 0$ \\ $\lambda_{1111} = 0$ \\ $m_H = 0$ \end{tabular} & \cellcolor{gray!18}\begin{tabular}[l]{@{}c@{}@{}}$m_{H^\pm} = 0$ \\ $m_H$ saddle \end{tabular} &  \cellcolor{gray!18}\begin{tabular}[l]{@{}c@{}@{}} $\mu_{12}^2=0$ \\ $\lambda_{1111}=0$ \\ $m_H = 0$ \\ $m_H$ degen. \end{tabular} & \cellcolor{gray!18}\begin{tabular}[l]{@{}c@{}@{}}$m_{H^\pm} = 0$ \\ $m_{H_{1,2}} = 0$ \end{tabular} \\ \hline

$(0, 0, v)$ & \cellcolor{gray!18}\begin{tabular}[l]{@{}c@{}@{}}$m_{H_{1,2}^\pm} = 0$ \\ $m_H$ saddle \end{tabular} & \cellcolor{gray!18}\begin{tabular}[l]{@{}c@{}@{}}$\lambda_{3333} = 0$ \\ $m_H = 0$\end{tabular} & \cellcolor{gray!18} \begin{tabular}[l]{@{}c@{}@{}} $m_H, m_{H^\pm}$ \\ saddle \end{tabular} & \cellcolor{gray!18} \begin{tabular}[l]{@{}c@{}@{}}$\lambda_{3333} = 0$ \\ $m_H = 0$\end{tabular} & \cellcolor{gray!18} \begin{tabular}[l]{@{}c@{}@{}}$m_H, m_{H^\pm}$ \\ saddle \end{tabular} & \cellcolor{gray!18} \begin{tabular}[l]{@{}c@{}@{}}$\lambda_{3333} = 0$ \\ $m_H = 0$\end{tabular} \\ \hline \hline

\end{tabular}
\end{center}
\end{table}}

{\renewcommand{\arraystretch}{1.4}
\begin{table}[H]
\caption{Classification of vacuum configurations for partial sign-flipping GCP-like GOOFy transformations. Notation follows Table~\ref{Table:Vacua_overall_X}.}
\label{Table:Vacua_diff_sgn_GCP}
\footnotesize
\begin{center}
\begin{tabular}{|c||c|c||c|c|} \hline\hline

& $\tilde{\mathbb{Z}}_2(\mathcal X^\prime)$ & $\tilde{\mathbb{Z}}_2(-\mathcal X^\prime)$ & $\tilde{\mathbb{Z}}_4(\mathcal X^\prime)$ & $\tilde{\mathbb{Z}}_4(-\mathcal X^\prime)$ \\ \hline \hline

$(v_1, v_2, v_3)$ & & $\lambda_{1332}^\mathrm{I} = \lambda_{1233}^\mathrm{I}$  & $\lambda_{1313}^\mathrm{I} = 0$  & $\lambda_{1313}^\mathrm{I} = 0$  \\ \hline \hline

$(v_1, v_2, 0)$ & \cellcolor{gray!18} \begin{tabular}[l]{@{}c@{}@{}} $\mu_{12}^2 = 0$ \\ $m_H = 0$ \end{tabular} & \cellcolor{gray!18} \begin{tabular}[l]{@{}c@{}@{}} $m_{H^\pm} = 0$ \\ $m_H \text{ saddle}$ \end{tabular} &  &  \\ \hline \hline

$(v_1, 0, v_3)$ & & $\lambda_{1332}^\mathrm{I} = \lambda_{1233}^\mathrm{I}$ & $\lambda_{1313}^\mathrm{I} = 0$ & $\lambda_{1313}^\mathrm{I} = 0$ \\ \hline

$(v_1, 0, 0)$ & \cellcolor{gray!18} \begin{tabular}[l]{@{}c@{}@{}} $\mu_{12}^2 = 0$ \\ $\lambda_{1111} = \lambda_{1112} = 0$ \\ $m_H = 0$\end{tabular} & \cellcolor{gray!18} \begin{tabular}[l]{@{}c@{}@{}} $m_{H^\pm} = 0$ \\ $m_H \text{ saddle}$ \end{tabular} &  & $\lambda_{1112} = 0$ \\ \hline

$(0, 0, v)$ &  \cellcolor{gray!18} \begin{tabular}[l]{@{}c@{}@{}} $m_{H^\pm}, m_H$ \\ saddle \end{tabular} & \cellcolor{gray!18} \begin{tabular}[l]{@{}c@{}@{}} $\lambda_{3333}=0$ \\ $m_H = 0$ \end{tabular}& \cellcolor{gray!18} \begin{tabular}[l]{@{}c@{}@{}} $m_{H^\pm}, m_H$ \\ saddle \end{tabular} & \cellcolor{gray!18} \begin{tabular}[l]{@{}c@{}@{}} $\lambda_{3333} = 0$ \\ $m_H = 0$ \end{tabular} \\ \hline \hline

\end{tabular}
\end{center}
\end{table}}

\subsection{Evaluating RG running}\label{Sec:Evaluating_RG}

Evaluating the RG stability of NHDMs requires determining whether the relations imposed on the scalar potential are preserved under radiative corrections. For symmetry-stabilised potentials, the vanishing of forbidden couplings is protected by 't~Hooft naturalness: since setting these couplings to zero enhances the symmetry of the theory, they define RG-invariant parameter subspaces. Nevertheless, the GOOFy potentials derived in the previous section must be examined explicitly, as the tree-level constraints need not be preserved beyond leading order. To verify their consistency under the RG flow, we employ \textsf{PyR@TE}~\cite{Sartore:2020gou} and \textsf{RGBeta}~\cite{Thomsen:2021ncy} codes. Throughout this work, we restrict the analysis to two-loop RG equations and neglect fermionic contributions (for a related analyses of the Yukawa sector under RGEs in 2HDMs, see Refs.~\cite{Ferreira:2023dke,Grzadkowski:2026gkx,Ferreira:2026ccg}).

Since the quartic sector is trivially invariant under a global sign flip ($\mathcal{X} = - \mathcal{I}$), stability depends on the quadratic terms. The simplest case, $V_2 = 0$, is straightforward: because the gauge, quartic, and Yukawa sectors cannot radiatively generate mass terms from a SI baseline, $V_2 = 0$ constitutes a stable UV fixed point in the presence of a regular HF or GCP symmetry.

For the non-trivial configurations with $V_2\neq 0$, we must impose $\mathrm{tr}(Y)=0$ (this condition was also observed in Ref.~\cite{Grzadkowski:2026gkx}), which is in fact necessary within the GOOFy framework. Indeed, due to the sign representation of $\mathcal X$, the invariance condition takes the form $\mathrm{tr}\left( \mathcal U^\dagger Y\,\mathcal U \right) = - \mathrm{tr}(Y)$; the non-zero eigenvalues of $Y$ must occur in $\pm$ pairs. 

The viability of this traceless condition depends entirely on how the transformation groups the scalar doublets into orbits:
\begin{itemize}
\item Single orbit: If a single generator connects all three doublets, unitarity dictates that their mass parameters must share the same absolute magnitude. Because an odd number of $SU(2)$ doublets cannot sum to zero using combinations of $\pm 1$, a completely degenerate traceless configuration is impossible in a 3HDM;

\item Two orbits ($SU(3)$ $\lambda_8$-like): If the transformation splits the fields into a doublet and a singlet, the potential takes the form $V_2 = \mu_{11}^2 (h_{11} + h_{22}) + \mu_{33}^2 h_{33}$. Because the symmetry treats these sub-representations as independent invariants, their $\beta$ functions run independently ($\beta_{\mu_{11}^2} \neq -2\beta_{\mu_{33}^2}$). Thus, a traceless condition imposed at tree level is radiatively destroyed;

\item Two orbits ($SU(3)$ $\lambda_3$-like): Aligning the $Y$ matrix with the $\lambda_3$ generator yields $V_2 = \mu_{11}^2 (h_{11} - h_{22})$. This structure preserves $\mu_{11}^2 + \mu_{22}^2 = 0$ along the entire RG trajectory, provided the quartic sector is governed by a permitted symmetry group. This maps perfectly onto the 2HDM traceless constraint given by eq.~\eqref{Eq:r0_2hdm_Couplings} and is the only solution from Subsection~\ref{Sec:Overall_X_discussion} that satisfies the representation mismatch criterion detailed in Subsection~\ref{Sec:Reps_mismatch}.

\end{itemize}

The situation changes for the partial sign-flipping GOOFy models discussed in Subsection~\ref{Sec:Partial_X_discussion}. To investigate their radiative stability, it is useful to first identify the maximal Abelian continuous symmetry of the scalar potential, namely $U(1)^{N-1}$. This symmetry corresponds to the maximal torus $T\subset SU(N)$, generated by the $N-1$ independent relative phase rotations of the scalar doublets after factoring out the overall $U(1)$ phase,
\begin{equation}
T = \left\lbrace \mathrm{diag}\left( e^{i \theta_1},\, e^{i \theta_2},\, \dots,\, e^{-i \sum_{k=1}^{N-1} \theta_k} \right) \;\middle|\; \theta_k \in \mathbb{R} \right\rbrace.
\end{equation}

Under $T$, the fundamental $\mathbf{N}$ of $SU(N)$ decomposes into $N$ one-dimensional representations. While $SU(N)$ singlets are invariant under every $\mathcal G\subset PSU(N)$, working with $T$ makes the full space of allowed field monomials explicit.

The most general scalar potential invariant under $U(1)^{N-1}$ may be written as
\begin{equation}\label{Eq:V_U1_general}
V_{U(1)^{N-1}} = \sum_{i=1}^N \mu_{ii}^2 h_{ii} + \sum_{i=1}^N \lambda_{iiii} h_{ii}^2 + \sum_{i<j}^N \lambda_{iijj} h_{ii}h_{jj} + \sum_{i<j}^N \lambda_{ijji} h_{ij}h_{ji}.
\end{equation}
The maximal-torus invariant terms cannot be arbitrarily omitted when applying either an HF or a GCP symmetry transformation: these symmetries may impose relations among their coefficients, but do not alter the underlying set of $U(1)^{N-1}$-invariant field monomials.

Within the bilinear formalism, these couplings map to the diagonal generalised Gell-Mann matrices. An inspection of the one-loop $\beta$ functions suggests that the $U(1)^{N-1}$-invariant sector is closed under the RG flow, whereas the phase-sensitive sector is not. While this observation is based on explicit multi-loop calculations rather than a general proof, it motivates the conjecture that the maximal-torus sector plays a critical role in the RG structure of arbitrary NHDMs.

From this perspective, the partial sign-flipping GOOFy models of Subsection~\ref{Sec:Partial_X_discussion} may be viewed as forcing some selected $U(1)^{N-1}$-allowed field monomials to zero—a condition that is, in general, not preserved by the RG flow. Consequently, the radiative stability of such constructions cannot be established solely by imposing \textit{ad hoc} parameter relations.

Even if the purely scalar contributions to a given $\beta$ function are arranged to cancel at a particular scale, this alignment is immediately destroyed by gauge contributions. The one-loop $\beta$ functions split into scalar, gauge, and mixed components. Forcing a total $\beta$ function to vanish in the absence of a symmetry would require the scalar quartic couplings to be fixed as non-trivial functions of the gauge parameters. Because gauge and quartic couplings obey mismatched RG scaling behaviours, such artificial relationships cannot survive along an extended RG trajectory. Our observations strongly imply that the partial sign-flipping models are not generically radiatively stable in the full theory.

Having established that partial sign-flipping GOOFy transformations are generically unstable under RG evolution, the purpose of the orbit construction developed in this work comes into sharper focus. It addresses a complementary question to recent studies of RG-stable relations in scalar field theories~\cite{Haber:2025cbb,deBoer:2026ktb,Pilaftsis:2026wyw}. Rather than searching for RG-invariant submanifolds or deriving sufficient conditions for their existence, the orbit method classifies the admissible scalar potentials themselves. The classification is entirely group-theoretical, formulated in terms of the decomposition of irreducible representations into sign-orbit pairs alongside the additional sign representation carried by $\mathcal{X}$.

From this perspective, the matrix $\mathcal X$ may be viewed as playing the role of a discrete spurion. Although it is not a dynamical field, it specifies how the conjugate sector transforms and therefore determines which invariant tensors are permitted. Different choices of $\mathcal X$ thus determine the admissible parameter sectors of the scalar potential, while the orbit decomposition records how the scalar multiplets populate the corresponding sign-orbit classes.

There is also a close connection with the complexification argument of Ref.~\cite{Haber:2025cbb}. There, an apparently accidental RG-stable relation, such as eq.~\eqref{Eq:r0_2hdm_Couplings}, is shown to arise as an ordinary symmetry after enlarging the field space. The extended GOOFy construction follows the converse logic. Rather than starting from an RG relation, one first enlarges the scalar representation space and then classifies all admissible symmetry embeddings through their orbit structure. In this sense, both approaches exploit an enlarged representation space, although they address complementary questions: Ref.~\cite{Haber:2025cbb} identifies RG-stable relations through symmetry enhancement, whereas the orbit method classifies which symmetry sectors can be realised in the first place.

The orbit construction also appears to be compatible with the outer automorphism criterion of Ref.~\cite{deBoer:2026ktb}. Consider $\mathbb Z_{2m}$ acting in the extended space as $\rho(g)= \mathrm{diag}(\mathcal U, -\mathcal U^\ast)$. The irreducible representations are organised into the sign-orbit pairs of eq.~\eqref{Eq:O_sign_pairs}. Consequently, the representation content is closed under the inversion automorphism $g\mapsto g^{-1}$, which exchanges the two members of each orbit pair. The orbit construction therefore organises the scalar potential into sectors compatible with the automorphism structure identified in Ref.~\cite{deBoer:2026ktb}, although it does not by itself establish RG stability.

Indeed, the logical roles of the two approaches are different. The orbit method determines which symmetry sectors are realisable and how they decompose into complementary orbit classes, while the outer automorphism and spurion approaches determine whether a given symmetry sector defines an RG-invariant hypersurface. The former is therefore a classification problem, whereas the latter concerns the behaviour of those classes under RG.

The discussion above raises the question of whether the observed correspondence extends beyond the 3HDM. Our construction should not be interpreted as a proof of RG stability. Rather, it identifies the symmetry-realisable orbit sectors and their associated eigenvalue structures. The existing RG analyses may then be viewed as providing an independent criterion for determining which of these sectors remain invariant under renormalisation.

With this motivation, it is instructive to examine bigger NHDMs. In the 4HDM the HF-like $\mathbb Z_{2m}$ realisations lead to the following eigenvalue structures:
\begin{equation}\label{Eq:4HDM_V2_HF_eigenvalues}
\begin{aligned}
\mathcal X = \mathrm{diag}(-1,\, -1,\, -1,\, -1) :{}& \quad \{\pm \lambda_1, 0, 0\}, \ \{\pm \lambda_1, \pm \lambda_2\}, \ \{0, 0, 0, 0\}, \\
\mathcal X = \mathrm{diag}(\phantom{-}1,\, -1,\, -1,\, -1)  :{}& \quad\{\lambda_1, 0, 0, 0\}, \ \{\pm \lambda_1, \lambda_2, 0\}, \\
\mathcal X = \mathrm{diag}(\phantom{-}1,\, \phantom{-}1,\, -1,\, -1) :{}& \quad \{\lambda_1, \lambda_2, 0, 0\}, \ \{\pm \lambda_1, \lambda_2, \lambda_3\}, \\
\mathcal X = \mathrm{diag}(\phantom{-}1,\, \phantom{-}1,\, \phantom{-}1,\, -1) :{}& \quad \{\lambda_1, \lambda_2, \lambda_3, 0\},
\end{aligned}
\end{equation}
where $\lambda_i$ denote the non-zero eigenvalues of the $Y$ matrix.

For the GCP-like realisations one obtains
\begin{equation}\label{Eq:4HDM_V2_GCP_eigenvalues}
\begin{aligned}
\mathcal X = \mathrm{diag}(-1,\, -1,\, -1,\, -1) :{}& \quad \{\pm \lambda_1, \pm \lambda_2\},\\
\mathcal X = \mathrm{diag}(\phantom{-}1,\, -1,\, -1,\, -1)  :{}& \quad \{\pm \lambda_1, \lambda_2, 0\}, \\
\mathcal X = \mathrm{diag}(\phantom{-}1,\, \phantom{-}1,\, -1,\, -1) :{}& \quad \{\pm \lambda_1, \lambda_2, \lambda_3\}, \\
\mathcal X = \mathrm{diag}(\phantom{-}1,\, \phantom{-}1,\, \phantom{-}1,\, -1) :{}& \quad \{\lambda_1, \lambda_2, \lambda_3, 0\}.
\end{aligned}
\end{equation}

Proceeding to the 5HDM (the number of orbit classes exceeds the number of available scalar doublets for $\mathbb{Z}_{12}$, while the maximal realisable symmetry is supposed to be $\mathbb{Z}_{16}$), the HF-like constructions give:
\begin{equation}\label{Eq:5HDM_V2_HF_eigenvalues}
\begin{aligned}
\mathcal X = \mathrm{diag}(-1,\, -1,\, -1,\, -1,\, -1) :{}& \quad \{\pm \lambda_1, 0, 0, 0\}, \ \{\pm \lambda_1, \pm \lambda_2, 0\}, \ \{0, 0, 0, 0, 0\}, \\
\mathcal X = \mathrm{diag}(\phantom{-}1,\, -1,\, -1,\, -1,\, -1)  :{}& \quad\{\pm \lambda_1, 0, 0, 0\}, \ \{\pm \lambda_1, \pm \lambda_2, 0\},\, \{0, 0, 0, 0, 0\} \\
\mathcal X = \mathrm{diag}(\phantom{-}1,\, \phantom{-}1,\, -1,\, -1,\, -1) :{}&
\hspace{6.5pt}\begin{array}{l}
\{\lambda_1,0,0,0,0\},\,\{\lambda_1,\lambda_2,0,0,0\},\\
\{\pm\lambda_1,\lambda_2,\lambda_3,0\},\,\{\pm\lambda_1,\lambda_2,0,0\},
\end{array}\\
\mathcal X = \mathrm{diag}(\phantom{-}1,\, \phantom{-}1,\, \phantom{-}1,\, -1,\, -1) :{}&
\hspace{6.5pt}\begin{array}{l}
\{\lambda_1, \lambda_2, 0, 0, 0\},\, \{\lambda_1, \lambda_2, \lambda_3, 0, 0\},\\
\{\pm \lambda_1, \lambda_2, \lambda_3, \lambda_4\},\, \{\pm \lambda_1, \lambda_2, \lambda_3, 0\}
\end{array}\\
\mathcal X = \mathrm{diag}(\phantom{-}1,\, \phantom{-}1,\, \phantom{-}1,\, \phantom{-}1,\, -1) :{}& \quad \{\lambda_1, \lambda_2, \lambda_3, 0, 0\},\, \{\lambda_1, \lambda_2, \lambda_3, \lambda_4, 0\},
\end{aligned}
\end{equation}
whereas the GCP-like realisations yield:
\begin{equation}\label{Eq:5HDM_V2_GCP_eigenvalues}
\begin{aligned}
\mathcal X = \mathrm{diag}(-1,\, -1,\, -1,\, -1,\, -1) :{}& \quad \{\pm \lambda_1, \pm \lambda_2, 0\}, \\
\mathcal X = \mathrm{diag}(\phantom{-}1,\, -1,\, -1,\, -1,\, -1)  :{}& \quad \{\pm \lambda_1, \pm \lambda_2, 0\}, \\
\mathcal X = \mathrm{diag}(\phantom{-}1,\, \phantom{-}1,\, -1,\, -1,\, -1) :{}& \quad \{\pm \lambda_1, \lambda_2, \lambda_3, 0\},\\
\mathcal X = \mathrm{diag}(\phantom{-}1,\, \phantom{-}1,\, \phantom{-}1,\, -1,\, -1) :{}& \quad \{\pm \lambda_1, \lambda_2, \lambda_3, \lambda_4\},\\
\mathcal X = \mathrm{diag}(\phantom{-}1,\, \phantom{-}1,\, \phantom{-}1,\, \phantom{-}1,\, -1) :{}& \quad \{\lambda_1, \lambda_2, \lambda_3, \lambda_4, 0\}.
\end{aligned}
\end{equation}

We have explicitly verified that only the cases with $\mathcal X = - \mathcal I$ are RG stable.

\section{Conclusions}

In this work, we have formalised the group-theoretical framework governing GOOFy transformations in the 3HDM. Treating these non-standard transformations as a representation-theoretic problem, we classified the admissible scalar sectors without asserting a definitive quantum-field-theoretic origin for the transformations themselves.  Crucially, we demonstrated that the apparent radiative stability of the isolated $r_0$ parameter in the 2HDM is an exceptional feature of the $SU(2)$ algebra. In higher $SU(N \ge 3)$ algebras, new radiative channels are introduced, obstructing a naive generalisation of the 2HDM construction.

A central result of this study is the representation mismatch criterion, which dictates the existence of a non-vanishing quadratic sector. We established that a GOOFy-compatible quadratic sector requires the symmetry group to admit a non-trivial one-dimensional sign character, equivalently, a non-trivial $\mathbb{Z}_2$ quotient of its Abelianisation, realised within the tensor product of the scalar representation and its conjugate. Consequently, distinct embeddings of the same symmetry group within the GOOFy framework can yield different potentials, a feature absent in standard HF and GCP frameworks.

To systematically organise these embeddings, we developed the sign-orbit construction for $\mathbb{Z}_{2m}$ groups. This technique reduces the identification of admissible bilinear sectors to a classification problem in character theory. Using this approach, we classified the surviving potentials for both overall and partial sign-flipping transformations.  However, algebraic consistency alone does not guarantee physical viability; the stationary-point equations in these restricted parameter spaces often lead to phenomenologically unacceptable massless or saddle-point vacuum states.

Furthermore, our RG analysis reveals a distinction between the two types of constructions. Overall sign-flipping transformations are generically RG-stable because they act uniformly as parity across the entire bilinear space. In contrast, partial sign-flipping transformations enforce arbitrary truncations on the scalar multiplet by forcing specific, $U(1)^{N-1}$-allowed (maximal-torus) parameters to zero. As we have established, this parameter subspace cannot be trivially truncated. We observe that these radiative effects inevitably destabilise partial sign-flipping configurations.

This framework provides a foundation for classification of larger GOOFy NHDMs.

\section*{Acknowledgements}

It is a pleasure to thank Andreas Trautner for enlightening discussions on properties of GOOFy transformations and pointing to several inconsistencies in the manuscript.

The work of AK was partially supported by the Spanish grants PID2023-147306NB-I00, CNS2024-154524 and CEX2023-001292-S (MICIU/AEI/10.13039/501100011033), as well as CIPROM/2021/054 (Generalitat Valenciana). AK would like to thank CFTP, Instituto Superior Técnico, University of Lisbon, for its hospitality.
IDMV acknowledges funding from Fundação para a Ciência e a Tecnologia (FCT) through the FCT Mobility program, and through the projects CFTP-FCT Unit UID/00777/2025 (\url{https://doi.org/10.54499/UID/00777/2025}), UIDB/FIS/00777/2020 and UIDP/FIS/00777/2020, \linebreak CERN/FIS-PAR/0019/2021, CERN/FIS-PAR/0002/2021, 2024.02004 CERN, which are partially funded through POCTI (FEDER), COMPETE, QREN and EU. IDMV thanks the University of Basel for hospitality.

\appendix

\section{The 3HDM in the bilinear formalism}\label{App:Bilinear_3HDM}

Here we present the 3HDM scalar potential in the bilinear formalism. To establish the basic notation, the nine complex $SU(2)$-invariant bilinears $h_{ij}$, $\{i,j\} \in \{1, 2, 3\}$ (strictly speaking not independent), can be decomposed into a single real singlet $r_0$ and a real vector $r_a$, with $a$ running from 1 to 8, using the Gell-Mann matrices $\lambda_a$:
\begin{equation}
h_{ij} = \frac{1}{\sqrt{3}} r_0 \delta_{ij} +  \sum_{a=1}^{8} r_a (\lambda_a)_{ij},
\end{equation}
where
\begin{subequations}
\begin{align}
r_0 ={}& \frac{1}{\sqrt{3}} \sum_{i=1}^{3} h_{ii},\\
r_a ={}& \frac{1}{2} \sum_{\{i,\,j\}=1}^{3} (\lambda_a)_{ij}\,  h_{ij}.
\end{align}
\end{subequations}

In this basis the scalar potential is given by:
\begin{equation}
V = M_0 r_0 + M_i r_i + \Lambda_{00} r_0^2 + \Lambda_{0i} r_0 r_i + \Lambda_{ij} r_i r_j,
\end{equation}

The vector components $M_a$ are defined by the relation:
\begin{equation}
M_a = - \frac{1}{2} \mathrm{tr}(Y \lambda_a).
\end{equation}

The quadratic part is given by:
\begin{equation}
M = \begin{pmatrix}
    -\frac{1}{\sqrt{6}} \left( \mu_{11}^2 + \mu_{22}^2 + \mu_{33}^2 \right) \\
    -\Re(\mu_{12}^2) \\
    \Im(\mu_{12}^2) \\
    \frac{1}{2} \left( \mu_{22}^2 - \mu_{11}^2 \right) \\
    -\Re(\mu_{13}^2) \\
    \Im(\mu_{13}^2) \\
    -\Re(\mu_{23}^2) \\
    \Im(\mu_{23}^2) \\
    -\frac{1}{2\sqrt{3}} \left( \mu_{11}^2 + \mu_{22}^2 - 2\mu_{33}^2 \right)
\end{pmatrix}^\mathrm{T}.
\end{equation}
Note that the $M_0$ element corresponds to the first line, while $M_8$ is the last element.

Because the quartic potential couples two $SU(2)$ bilinear singlets together, one must project over two sets of Gell-Mann matrices:
\begin{equation}
\Lambda_{ab} = \frac{1}{4} \sum_{i,j,k,l} Z_{ijkl} (\lambda_a)_{ji} (\lambda_b)_{lk}.
\end{equation}

The quadratic part is given by:
\begin{align}
\Lambda_{00} ={}& \frac{1}{3} \left( \lambda_{1111} + \lambda_{2222}  + \lambda_{3333} + \lambda_{1122} + \lambda_{1133}  + \lambda_{2233} \right),\\
\Lambda_{0i} ={}& \begin{pmatrix}
    -\frac{1}{\sqrt{3}} \Im \big( \lambda_{1112} + \lambda_{1222} + \lambda_{1233} \big) \\
    \phantom{-}\frac{1}{2\sqrt{3}} \big( 2\lambda_{1111} + \lambda_{1133} - 2\lambda_{2222} - \lambda_{2233} \big) \\
    \phantom{-}\frac{1}{\sqrt{3}} \Re \big( \lambda_{1113} + \lambda_{1322} + \lambda_{1333} \big) \\
    -\frac{1}{\sqrt{3}} \Im \big( \lambda_{1113} + \lambda_{1322} + \lambda_{1333} \big) \\
    \phantom{-}\frac{1}{\sqrt{3}} \Re \big( \lambda_{1123} + \lambda_{2223} + \lambda_{2333} \big) \\
    -\frac{1}{\sqrt{3}} \Im \big( \lambda_{1123} + \lambda_{2223} + \lambda_{2333} \big) \\[6pt]
   \phantom{-}\frac{1}{6} \big( 2\lambda_{1111} + 2\lambda_{1122} - \lambda_{1133} + 2\lambda_{2222} - \lambda_{2233} - 4\lambda_{3333} \big)
\end{pmatrix}^\mathrm{T},
\end{align} 
where
\begin{equation}
\begin{aligned}[t]
    \Lambda_{11} ={}& 2\Re(\lambda_{1212}) + \lambda_{1221}, \\
    \Lambda_{12} ={}& -2\Im(\lambda_{1212}), \\
    \Lambda_{13} ={}& \Re \big( \lambda_{1112} - \lambda_{1222} \big), \\
    \Lambda_{14} ={}& \Re \big( \lambda_{1213} + \lambda_{1321} \big), \\
    \Lambda_{15} ={}& -\Im \big( \lambda_{1213} + \lambda_{1321} \big), \\
    \Lambda_{16} ={}& \Re \big( \lambda_{1223} + \lambda_{2123} \big), \\
    \Lambda_{17} ={}& -\Im \big( \lambda_{1223} + \lambda_{2123} \big), \\
    \Lambda_{18} ={}& \frac{1}{\sqrt{3}} \Re \big( \lambda_{1112} + \lambda_{1222} - 2\lambda_{1233} \big),\\
    \Lambda_{22} ={}& -2\Re(\lambda_{1212}) + \lambda_{1221}, \\
    \Lambda_{23} ={}& -\Im \big( \lambda_{1112} - \lambda_{1222} \big), \\
    \Lambda_{24} ={}& -\Im \big( \lambda_{1213} - \lambda_{1321} \big), \\
    \Lambda_{25} ={}& -\Re \big( \lambda_{1213} - \lambda_{1321} \big), \\
    \Lambda_{26} ={}& -\Im \big( \lambda_{1223} - \lambda_{2123} \big), \\
    \Lambda_{27} ={}& -\Re \big( \lambda_{1223} - \lambda_{2123} \big), \\
    \Lambda_{28} ={}& -\frac{1}{\sqrt{3}} \Im \big( \lambda_{1112} + \lambda_{1222} - 2\lambda_{1233} \big),\\
    \Lambda_{33} ={}& \lambda_{1111} - \lambda_{1122} + \lambda_{2222}, \\
    \Lambda_{34} ={}& \Re \big( \lambda_{1113} - \lambda_{1322} \big), \\
    \Lambda_{35} ={}& -\Im \big( \lambda_{1113} - \lambda_{1322} \big), \\
    \Lambda_{36} ={}& \Re \big( \lambda_{1123} - \lambda_{2223} \big), \\
    \Lambda_{37} ={}& -\Im \big( \lambda_{1123} - \lambda_{2223} \big),
\end{aligned} 
\qquad\qquad
\begin{aligned}[t]
    \Lambda_{38} ={}& \frac{1}{\sqrt{3}} \big( \lambda_{1111} - \lambda_{1133} - \lambda_{2222} + \lambda_{2233} \big),\\
    \Lambda_{44} ={}& 2\Re(\lambda_{1313}) + \lambda_{1331}, \\
    \Lambda_{45} ={}& -2\Im(\lambda_{1313}), \\
    \Lambda_{46} ={}& \Re \big( \lambda_{1323} + \lambda_{2331} \big), \\
    \Lambda_{47} ={}& -\Im \big( \lambda_{1323} + \lambda_{2331} \big), \\
    \Lambda_{48} ={}& \frac{1}{\sqrt{3}} \Re \big( \lambda_{1113} + \lambda_{1322} - 2\lambda_{1333} \big),\\
    \Lambda_{55} ={}& -2\Re(\lambda_{1313}) + \lambda_{1331},\\
    \Lambda_{56} ={}& -\Im \big( \lambda_{1323} - \lambda_{2331} \big), \\
    \Lambda_{57} ={}& -\Re \big( \lambda_{1323} - \lambda_{2331} \big), \\
    \Lambda_{58} ={}& -\frac{1}{\sqrt{3}} \Im \big( \lambda_{1113} + \lambda_{1322} - 2\lambda_{1333} \big),\\
    \Lambda_{66} ={}& 2\Re(\lambda_{2323}) + \lambda_{2332}, \\
    \Lambda_{67} ={}& -2\Im(\lambda_{2323}), \\
    \Lambda_{68} ={}& \frac{1}{\sqrt{3}} \Re \big( \lambda_{1123} + \lambda_{2223} - 2\lambda_{2333} \big),\\
    \Lambda_{77} ={}& -2\Re(\lambda_{2323}) + \lambda_{2332}, \\
    \Lambda_{78} ={}& -\frac{1}{\sqrt{3}} \Im \big( \lambda_{1123} + \lambda_{2223} - 2\lambda_{2333} \big),\\
    \Lambda_{88} ={}& \frac{1}{3} \big( \lambda_{1111} + \lambda_{1122}- 2\lambda_{1133}\\
                    & \quad  + \lambda_{2222} - 2\lambda_{2233} + 4\lambda_{3333} \big).
\end{aligned}
\end{equation}

\section{Some examples of the GOOFy potentials}\label{App:V2_GOOFy_options}

We fix $\mathcal X$ to be diagonal, and it can only take values of $\pm 1$. This specific basis choice simplifies the systematic classification of GOOFy transformations. In this frame, a set of discrete signatures defined by the combinations of $\pm 1$ on the diagonal can be identified:
\begin{itemize}
    \item $\mathcal{X} = \mathcal{I}$: The standard symmetry transformations are recovered in this case;
    \item $\mathcal{X} = -\mathcal{I}$: Introduces a global sign flip;
    \item One anomalous direction: $\text{diag}(-1, 1, 1)$ and permutations;
    \item Two anomalous directions: $\text{diag}(-1, -1, 1)$ and permutations.
\end{itemize}

To explicitly demonstrate how the scalar potential depends simultaneously on  $\mathcal{U}$ and $\mathcal{X}$, we evaluate the constraints
\begin{subequations}
\begin{align}
\text{HF-like case: }& Y\phantom{^\ast} = \mathcal{X} \mathcal{U}^\dagger Y \mathcal{U}, \label{Eq:Y_to_hat_X_HF}\\
\text{GCP-like case: }& Y^\ast = \mathcal{X} \mathcal{U}^\dagger Y \mathcal{U}, \label{Eq:Y_to_hat_X_GCP}
\end{align}
\end{subequations}
across distinct embeddings of a symmetry $\mathcal{U}$.

\subsection{The trivial embedding}\label{App:V2_discussion}

First of all consider $\mathcal{U} = \mathcal{I}$ implemented as an HF-like transformation. The constraint of eq.~\eqref{Eq:Y_to_hat_X_HF} simplifies to $Y = \mathcal{X} Y$. A choice of $\mathcal{X}_A=\mathrm{diag}(-1, -1, 1)$ yields:
\begin{equation}
\begin{pmatrix} \mu_{11}^2 & \mu_{12}^2 & \mu_{13}^2 \\ (\mu_{12}^2)^\ast & \mu_{22}^2 & \mu_{23}^2 \\ (\mu_{13}^2)^\ast & (\mu_{23}^2)^\ast & \mu_{33}^2 \end{pmatrix} 
\to \begin{pmatrix} -\mu_{11}^2 & -\mu_{12}^2 & -\mu_{13}^2 \\ -(\mu_{12}^2)^\ast & -\mu_{22}^2 & -\mu_{23}^2 \\ (\mu_{13}^2)^\ast & (\mu_{23}^2)^\ast & \mu_{33}^2 \end{pmatrix}.
\end{equation}
Demanding invariance requires the matrix to map to itself, forcing the upper $2 \times 3$ block to zero. The $\mathcal{X}_A$ matrix acts as a projector, preserving only the subspace aligned with its $+1$ eigenvalue:
\begin{equation}
V_2 = \mu_{33}^2 h_{33}.
\end{equation}

In a similar manner, evaluating the transformation under $\mathcal{X}_B = \operatorname{diag}(1, 1, -1)$ yields:
\begin{equation}
\begin{pmatrix} \mu_{11}^2 & \mu_{12}^2 & \mu_{13}^2 \\ (\mu_{12}^2)^\ast & \mu_{22}^2 & \mu_{23}^2 \\ (\mu_{13}^2)^\ast & (\mu_{23}^2)^\ast & \mu_{33}^2 \end{pmatrix} 
\to \begin{pmatrix} \mu_{11}^2 & \mu_{12}^2 & \mu_{13}^2 \\ (\mu_{12}^2)^\ast & \mu_{22}^2 & \mu_{23}^2 \\ -(\mu_{13}^2)^\ast & -(\mu_{23}^2)^\ast & -\mu_{33}^2 \end{pmatrix}.
\end{equation}
The invariant quadratic potential is given by:
\begin{equation}
V_2 = \mu_{11}^2 h_{11} + \mu_{22}^2 h_{22} + \left(\mu_{12}^2 h_{12} + \mathrm{h.c.}\right).
\end{equation}

Remarkably, whilst the quadratic potentials are different, the quartic scalar potentials for both cases coincide. Because the kinetic anomaly acts asymmetrically, the invariance condition evaluates to $\lambda_{ijkl} = \mathcal{X}_{ii}\mathcal{X}_{kk} \lambda_{ijkl}$. However, to maintain a Hermitian scalar potential, the conjugate interaction ($h_{ji} h_{lk}$) must independently satisfy this same rule, imposing a simultaneous constraint on the remaining indices. Thus, a quartic parameter is permitted to survive if and only if it strictly satisfies the dual condition:
\begin{equation}
\lambda_{ijkl} = \mathcal{X}_{ii}\mathcal{X}_{kk} \lambda_{ijkl} \quad \text{and} \quad \lambda_{ijkl} = \mathcal{X}_{jj}\mathcal{X}_{ll} \lambda_{ijkl}.
\end{equation}

Because $\mathcal{X}_A = -\mathcal{X}_B$, the product is identical in both cases, $\mathcal{X}_{A,ii}\, \mathcal{X}_{A,kk} = \mathcal{X}_{B,ii}\, \mathcal{X}_{B,kk}$. Consequently, the surviving parameter space is identical for both $\mathcal{X}$. The resulting quartic potential is given by
\begin{equation}\label{V4_Z2_X_AB}
\begin{aligned}
V_4 ={}& \lambda_{1111} h_{11}^2 + \lambda_{2222} h_{22}^2 + \lambda_{3333} h_{33}^2 + \lambda_{1221} h_{12} h_{21} + \lambda_{1122}h_{11}h_{22} \\
&+ \bigg\{ \lambda_{1112} h_{11}h_{12} + \lambda_{1222} h_{12}h_{22} + \lambda_{1323} h_{13}h_{23} \\
&\qquad+ \lambda_{1212} h_{12}^2 + \lambda_{1313} h_{13}^2 + \lambda_{2323} h_{23}^2 + \mathrm{h.c.} \bigg\}.
\end{aligned}
\end{equation}

This structure resembles the standard $\mathbb{Z}_2$-invariant model generated by $\operatorname{diag}(1, 1, -1)$, but with a crucial distinction. Because a parameter survives only if both the primary and conjugate index pairs satisfy the anomaly condition, any cross-coupling that mixes the eigenspaces is strictly removed. Thus, whilst the pure self-interaction $h_{33}^2$ survives ($\mathcal{X}_{33}\mathcal{X}_{33} = +1$), any mixed terms connecting the third generation to the first two, such as the diagonal cross-coupling $h_{11}h_{33}$ or the misaligned conjugate pair $h_{13}h_{31}$, are forbidden.

It is crucial to observe that the surviving terms involving $\{ h_1,\, h_2\}$ construct an embedded, general 2HDM. Because any $U(2)$ basis transformation acting on the $\{h_1,\, h_2\}$ subspace commutes with the topological background, the GOOFy constraints are blind to such rotations. Consequently, the above potential contains three redundant degrees of freedom. One can exploit this residual basis freedom to simplify the potential.

In the GCP case evaluating $\mathcal{X}_A$ yields:
\begin{equation}
\begin{pmatrix} \mu_{11}^2 & (\mu_{12}^2)^\ast & (\mu_{13}^2)^\ast \\ \mu_{12}^2 & \mu_{22}^2 & (\mu_{23}^2)^\ast \\ \mu_{13}^2 & \mu_{23}^2 & \mu_{33}^2 \end{pmatrix} 
\to \begin{pmatrix} -\mu_{11}^2 & -\mu_{12}^2 & -\mu_{13}^2 \\ -(\mu_{12}^2)^\ast & -\mu_{22}^2 & -\mu_{23}^2 \\ (\mu_{13}^2)^\ast & (\mu_{23}^2)^\ast & \mu_{33}^2 \end{pmatrix}.
\end{equation}
The surviving quadratic potential is:
\begin{equation}
V_2 = \mu_{33}^2 h_{33} - 2(\mu_{12}^2)^\mathrm{I} \Im(h_{12}).
\end{equation}

Now, evaluating $\mathcal{X}_B$ under GCP yields:
\begin{equation}
\begin{pmatrix} \mu_{11}^2 & (\mu_{12}^2)^\ast & (\mu_{13}^2)^\ast \\ \mu_{12}^2 & \mu_{22}^2 & (\mu_{23}^2)^\ast \\ \mu_{13}^2 & \mu_{23}^2 & \mu_{33}^2 \end{pmatrix} 
\to\begin{pmatrix} \mu_{11}^2 & \mu_{12}^2 & \mu_{13}^2 \\ (\mu_{12}^2)^\ast & \mu_{22}^2 & \mu_{23}^2 \\ -(\mu_{13}^2)^\ast & -(\mu_{23}^2)^\ast & -\mu_{33}^2 \end{pmatrix}.
\end{equation}
The invariant scalar potential is therefore given by:
\begin{equation}
V_2 = \mu_{11}^2 h_{11} + \mu_{22}^2 h_{22} + 2(\mu_{12}^2)^\mathrm{R} \Re(h_{12}).
\end{equation}

For completeness, we note the limits of the $\mathcal{X} = -\mathcal{I}$ case. The sign flip forces $Y = -Y$, which results in $V_2 = 0$, enforcing SI in the mass sector. However, because the constraint on the quartic tensor requires two insertions, $\mathcal{X}^2 = \mathcal{I}$, the minus signs compensate one another. Consequently, the quartic potential $V_4$ remains entirely transparent to the transformation, yielding the standard couplings of the most general 3HDM.

\subsection[The \texorpdfstring{$\mathbb{Z}_2$}{Z2} embeddings]{The $\mathbb{Z}_2$ embeddings}

Consider the $\mathbb{Z}_2$ symmetry generated by $\mathcal{U} = \mathrm{diag}(1, 1, -1)$. Because both $\mathcal{U}$ and $\mathcal{X}$ are diagonal in this frame, we can define a composite operator $S_{ij} = \mathcal{X}_{ii}\, \mathcal{U}_{ii}^\ast\, \mathcal{U}_{jj}$ which describes the net transformation of the bilinear $h_{ij}$. Explicitly:
\begin{align}
\mu^2_{ij} ={}& S_{ij} \, \mu^2_{ij}, \\
\lambda_{ijkl} ={}& S_{ij} S_{kl} \, \lambda_{ijkl}.
\end{align}
Note that off-diagonal terms must also satisfy Hermiticity, requiring $S_{ji} = S_{ij}$.

As it turns out, this case produces the exact same parameter space as the trivial identity embedding $\mathcal{U} = \mathcal{I}$: $\mathcal{X}_A (-\mathcal{X}_A)^\dagger Y (-\mathcal{X}_A) =  \mathcal{X}_A$ and $\mathcal{X}_B (\mathcal{X}_B)^\dagger Y (\mathcal{X}_B) = Y \mathcal{X}_B$, note that $\mathcal{U} = \mathcal{X}_B = - \mathcal{X}_A$.

To fully grasp the constraints of the GOOFy framework, it is instructive to fix the generator $\mathcal{U}$ and consider other diagonal permutations of $\mathcal{X}$. The action of $\mathcal{U}$ is defined by its unique $-1$ eigenspace, pointing along the $h_3 \mapsto -h_3$ direction. If we apply a collinear transformation, $\mathcal{X} = \pm\, \mathcal{U}$, the eigenspaces are parallel. In this limit, the surviving quadratic terms correspond simply to the $+1$ eigenspace of $\mathcal{X}$ alone.

However, a shift occurs if we permute the anomaly matrix, generating a mismatch. Suppose we apply $\mathcal{X} = \mathrm{diag}(-1, 1, -1)$. This anomaly places a $-1$ on $h_1$ but a $+1$ on $h_2$, fracturing the degenerate $\{h_1,\, h_2\}$ plane of the symmetry generator.

This fracturing has direct consequences for the scalar potential:
\begin{itemize}
\item Diagonal masses in the anomalously flipped directions map to their own negatives and vanish entirely.

\item The $\mathcal X$ matrix can selectively compensate for the phase shift induced by the underlying transformation. For example, the $\mathcal{X}: h_1 \sim-1$  cancels the relative $-1$ phase shift between $h_1$ and $h_3$. This preserves the $h_{13}$ cross-mixing, whilst the uncompensated $h_{23}$ mixing is not allowed.
\end{itemize}

To be more explicit, we list different potentials for the choice of $\mathcal{U} = \mathrm{diag}(1, 1, -1)$:
\begin{subequations}
\begin{align}
V\left(\mathcal{X} = \mathrm{diag}(-1, -1, -1)\right) &= \{ \mu_{13}^2 h_{13} + \mu_{23}^2 h_{23} + \mathrm{h.c.} \} + V_4, \\
V\left(\mathcal{X} = \mathrm{diag}(-1, -1, \phantom{-}1)\right) &= \mu_{33}^2 h_{33} + V_4', \\
V\left(\mathcal{X} = \mathrm{diag}(\phantom{-}1, \phantom{-}1, -1)\right) &= \mu_{11}^2 h_{11} + \mu_{22}^2 h_{22} + \{ \mu_{12}^2 h_{12} + \mathrm{h.c.} \} + V_4', \\
V\left(\mathcal{X} = \mathrm{diag}(-1, \phantom{-}1, \phantom{-}1)\right) &= \mu_{22}^2 h_{22} + \mu_{33}^2 h_{33} + V_4'', \\
V\left(\mathcal{X} = \mathrm{diag}(\phantom{-}1, -1, -1)\right) &= \mu_{11}^2 h_{11} + \{ \mu_{23}^2 h_{23} + \mathrm{h.c.} \} + V_4''.
\end{align}
\end{subequations}
Here, $V_4$ denotes a $\mathbb{Z}_2$-symmetric potential, the $V_4'$ potential was presented in eq.~\eqref{V4_Z2_X_AB}, and $V_4''$ is given by
\begin{equation}
\begin{aligned}
V_4 ={}& \lambda_{1111} h_{11}^2 + \lambda_{2222} h_{22}^2 + \lambda_{3333} h_{33}^2 + \lambda_{2332} h_{23} h_{32} + \lambda_{2233}h_{22}h_{33}\\
& + \left\lbrace \lambda_{1321} h_{13}h_{21} + \lambda_{1123} h_{11}h_{23} + \lambda_{1212} h_{12}^2 + \lambda_{1313} h_{13}^2 + \lambda_{2323} h_{23}^2 + \mathrm{h.c.} \right\rbrace.
\end{aligned}
\end{equation}

\subsection{Interpretation in the bilinear formalism}

The structure of the admissible quadratic sector can also be understood directly within the bilinear formalism. Since the quadratic potential is given by
\begin{equation}
V_2 = M_\mu r_\mu,
\end{equation}
the problem reduces to identifying which bilinear directions remain invariant under a given transformation. 

To systematise this, one may define the transformation shift
\begin{equation}
\Delta r_\mu \equiv r_\mu^\prime - r_\mu.
\end{equation}
The condition $\Delta r_\mu = 0$ thereby isolates the invariant subspace that contributes to $V_2$.

For the off-diagonal terms interpretation is immediate: if $\Delta r_\mu = 0$ for a phase-sensitive direction, the corresponding $h_{ij}$ term survives in $V_2$. The diagonal sector is more subtle. The bilinears $r_0$, $r_3$, and $r_8$ span the $U(1)^2$ maximal torus. Because they are constructed as linear combinations of the individual diagonal quantities $h_{ii}$, the condition $\Delta r_\mu = 0$ for $\mu \in \{0,3,8\}$ does not directly isolate individual $h_{ii}$ operators. Instead, one must explicitly invert the basis transformation to reconstruct the surviving combinations, determining exactly which diagonal field terms remain present in the potential.

\bibliography{ref}

\end{document}